\documentclass[aps,preprint,onecolumn,floatfix,superscriptaddress,nofootinbib]{revtex4-1}
\usepackage[T1]{fontenc} 

\newcommand{\sv}{\ensuremath{\langle\sigma v\rangle}}
\newcommand{\dnde}{\ensuremath{\frac{dN_e}{dE}}}

\newcommand{\mev}{\ensuremath{\,\mathrm{MeV}}}
\newcommand{\gev}{\ensuremath{\,\mathrm{GeV}}}
\newcommand{\tev}{\ensuremath{\,\mathrm{TeV}}}

\usepackage{amsmath}
\usepackage{graphicx,bm}
\usepackage[colorlinks,citecolor=blue]{hyperref}
\usepackage[caption=false]{subfig}
\usepackage{slashed}
\begin{document}
\title{Sensitivity of SKA to dark matter induced radio emission}

\author{Zhanfang Chen}
\affiliation{Key Laboratory of Dark Matter and Space Astronomy, Purple Mountain Observatory, Chinese Academy of Sciences, Nanjing 210033, China}
\affiliation{School of Astronomy and Space Science, University of Science and Technology of China, Hefei, Anhui 230026, China}

\author{Yue-Lin Sming Tsai}
\email{smingtsai@pmo.ac.cn}
\affiliation{Key Laboratory of Dark Matter and Space Astronomy, Purple Mountain Observatory, Chinese Academy of Sciences, Nanjing 210033, China}

\author{Qiang Yuan}
\affiliation{Key Laboratory of Dark Matter and Space Astronomy, Purple Mountain Observatory, Chinese Academy of Sciences, Nanjing 210033, China}
\affiliation{School of Astronomy and Space Science, University of Science and Technology of China, Hefei, Anhui 230026, China}

\begin{abstract}
Conventionally, one can constrain the dark matter (DM) interaction 
with DM mass heavier than GeV by searching for DM induced synchrotron emission 
in the radio frequency band. 
However, an MeV DM can also generate detectable radio emission if electrons and positrons 
produced by DM annihilation or decay undergoes inverse Compton scattering (ICS) with 
the cosmic microwave background. The upcoming radio telescope Square Kilometre Array (SKA) 
is designed to operate with extremely high sensitivity. 
We investigate the capability of the SKA to detect DM particles in a board mass range
from MeV to TeV, for both annihilation and decay scenarios.   
In this paper, we consider the sensitivities of the future SKA first and second phase (SKA1 and SKA2). 
As a comprehensive study, we systematically study the impacts on the DM-induced signal 
computation from the magnetic field strengths and particle diffusion coefficients.  
We compare the detection potential of four very different sources: 
two dwarf spheroidal galaxies (Draco and Segue 1), one radio-poor cluster (A2199), 
and one DM-rich ultra-diffuse galaxy (Dragonfly 44). 
We project the SKA1 and SKA2 sensitivities with the exposure of 100 hours
on the annihilation cross section and decay time for DM mass from MeV to TeV 
by considering two different leptonic final states $e^+ e^-$ and $\mu^+\mu^-$. 
\end{abstract}

\date{\today}

\maketitle

\section{introduction} 
The anisotropies in the cosmic microwave background show that around a quarter of
the Universe consists of dark matter (DM). However, 
the existence of DM is only revealed by its gravitational 
interaction with visible matter. 
Nevertheless, people do not yet discard the assumption that 
DM might interact with visible matter via non-gravitational interactions.
After decades of efforts to search for non-gravitational interactions 
between DM and visible matter, significant progress has been reached, 
such as the underground DM direct detection experiments~\cite{Liu:2017drf, Schumann:2019eaa}, colliders~\cite{Buchmueller:2017qhf, Kahlhoefer:2017dnp}, or 
photon and cosmic ray observations known as DM indirect detection~\cite{Leane:2020liq, Gaskins:2016cha,PerezdelosHeros:2020qyt}. 
The sensitivities of these measurements have been improved rapidly in recent years, 
especially for the DM mass greater than $10\gev$. 
Therefore, the direction of the future DM searches of non-gravitational interactions 
should aim for either the lighter DM mass region or the extremely weak coupling region. 
These regions require detectors with a smaller threshold energy or larger exposure.

Compared with collider and direct detection, the strategy of DM indirect detection 
can be more flexible by looking for different targets or using various telescopes 
with very different threshold energy.  
Taking the $\gamma$-ray telescope as an example, 
the Fermi Gamma-ray Space Telescope has reported 
the current best limits of DM annihilation cross section ($b\bar{b}$ final state) 
for DM mass around $60\gev$~\cite{Fermi-LAT:2016uux} 
based on a combined analysis of $15$ dwarf galaxies (dSphs).  
The strongest constraint in the TeV mass range is given by the ground-based 
telescopes HESS through the observation of the Galactic Center~\cite{Abdallah:2016ygi}. 
The future ground-based telescopes Cherenkov Telescope Array (CTA)~\cite{Silverwood:2014yza}
can significantly improve the sensitivity and probe 
the thermal DM cross section $\sv\simeq 10^{-26}$~cm$^{3} s^{-1}$ 
for DM mass between $200\gev$ and $10\tev$.

If the standard model electrons and positrons are produced by DM annihilation or decay,   
they may lose their energies either by inverse Compton scattering (ICS) with the 
Cosmic microwave background (CMB) photons or synchrotron radiation from 
gyration in the magnetic field, and radiate in the radio frequencies. 
Both ICS and synchrotron emission can contribute to this frequency, 
depending on the DM mass. 
If DM mass is heavier than $\gev$, the synchrotron emission can be detected 
by current or future radio telescopes~\cite{Colafrancesco:2005ji,Spekkens:2013ik,McDaniel:2017ppt}. 
For DM mass smaller than $100\mev$, the peak frequency of the DM synchrotron 
emission is too low to be detected.
However, one can still observe DM ICS within the radio frequency range~\cite{Dutta:2020lqc}.
A recent deep radio observation of the dSph Reticulum II with the Australia Telescope Compact 
Array shows a competitive limit for DM masses between $10\gev$ to $100\gev$~\cite{Regis:2017oet}.
Similar studies are also performed by Ref.~\cite{Natarajan:2015hma} with the Green Bank Telescope, 
by Ref.~\cite{Kar:2019hnj} with the combined data from the Murchison Widefield Array and the 
Giant Metre-wave Radio Telescope, and by Ref.~\cite{Basu:2021zfg} with a stacking analysis of 
23 dSph galaxies using data from TIFR GMRT Sky Survey.

The Square Kilometre Array (SKA) is a planned radio telescope array, 
which will be built in Australia and South Africa. The first phase of
SKA, known as SKA1, is going to start its construction in 2021.
The frequency coverage of the SKA is from 50~MHz to 50~GHz. 
The outstanding resolution and sensitivity of the SKA could effectively probe 
GeV scale DM via synchrotron emission~\cite{Kar:2019cqo,Cembranos:2019noa}. 
A recent study~\cite{Beck:2021xsv} also shows that the multi-lepton anomalies 
at the LHC~\cite{vonBuddenbrock:2017gvy} may be probed by future observations 
of Reticulum II. Some recent studies~\cite{Kar:2019cqo,Cembranos:2019noa} have 
estimated that the SKA sensitivity on the annihilation cross section of DM 
with masses heavier than GeV.  For DM mass range below GeV, the ICS emission
is shown to be more relevant than the synchrotron emission \cite{Dutta:2020lqc}. 
The SKA sensitivity is also discussed in probing the DM induced synchrotron radiation from 
the DM-rich Ultra-faint Dwarf Galaxies~\cite{Bhattacharjee:2020phk}.
In this work, we estimate the SKA sensitivities on DM annihilation cross 
section and decay rate in a wide mass range from $\mev$ to $\tev$. 
Both the ICS and synchrotron emission caused by either DM annihilation or decay
are considered. We study comprehensively the impacts of the poorly unknown 
magnetic field strengths and diffusion coefficients. We also discuss the 
detectability based on four different sources (Segue1, Draco, A2199, and Dragonfly 44) 
with different properties.

The SKA1 sensitivity has been investigated   
for both the DM induced synchrotron~\cite{Colafrancesco:2014coa,Kar:2018rlm,
Kar:2019cqo,Cembranos:2019noa,Ghosh:2020ipv,Beck:2021xsv} 
and ICS~\cite{Dutta:2020lqc} in the literature, 
but it can be more useful to consistently consider them together from MeV to TeV. 
For example, the present works of SKA sensitivity for DM induced synchrotron  
are mainly focusing on DM mass greater than $5\gev$~\cite{Colafrancesco:2014coa,Kar:2018rlm,
Kar:2019cqo,Cembranos:2019noa,Ghosh:2020ipv,Beck:2021xsv}. 
On the other hand, the SKA sensitivity for the DM induced ICS is only presented 
based on DM mass less than $100\mev$~\cite{Dutta:2020lqc} for  
DM annihilation and decay to $e^+ e^-$ final state. 
Therefore, our work fills the mass gap between $100\mev$ to $5\gev$ 
and we compute both synchrotron and ICS emission with a consistent model configuration.  
Our work shows that the synchrotron contribution is equally important as  
ICS one in this mass gap.  
Comparing with the DM ICS limits in Ref.~\cite{Dutta:2020lqc}, 
we include not only $e^+ e^-$ final state but also the $\mu^+\mu^-$ channel.

The paper is organized as follows. In Sec.~\ref{sec:SKA}, the SKA sensitivity 
is briefly described. We discuss the two phases of SKA1 and SKA2. 
In Sec.~\ref{sec:radio}, we recap the derivation of radio SED and show its 
variation when changing the magnetic field strength and diffusion coefficient. 
In the last part of Sec.~\ref{sec:radio}, we also survey using different sources. 
In Sec.~\ref{sec:result}, we present the upper limits of DM annihilation cross section and 
DM decay rate based on SKA1 and SKA2 sensitivity, assuming no radio emission is detected. 
We summarize our results in Sec.~\ref{sec:conclusion}.

\section{The Square Kilometre Array}
\label{sec:SKA}
The SKA is designed to detect radio in the range between 50~MHz to 50~GHz. 
The strongest strength of the SKA lies in its high sensitivity and energy resolution.     
Particularly, the detection of dSphs will benefit from the SKA strength 
because dSphs usually are faint radio sources.  
Some recently developed telescopes such as the Green Bank Telescope (GBT) 
and the Australia Telescope Compact Array (ATCA) make radio observations of 
several dSphs~\cite{Kar:2019hnj,Regis:2017oet,Natarajan:2015hma}. 
In these studies, null-detection have been observed and their sensitivities of radio emission 
is $\approx\mathcal{O}({\rm mJy})$ ($\sim 10^{-26}$ erg$\cdot$cm$^2\cdot s^{-1}\cdot$Hz$^{-1}$). 
In the near future, 
we expect that the SKA array would detect radio emission with 
the sensitivities of $\mathcal{O}(\mu {\rm Jy})$~\cite{Dewdney:2013ard}.

The minimum flux detected by the SKA telescope can be obtained by the following equation~\cite{Cembranos:2019noa}: 
\begin{eqnarray}
 S_{\text{min}}=\frac{2 k_{b}S_{D} T_{\text{sys}}}{\eta_{s}A_{e}(\eta_{\text{pol}}\mathcal{T}\Delta\nu)^{1/2}}\,,
\label{S_min}
\end{eqnarray}
where $k_{b}$ represents the Boltzmann constant, $\eta_{s} $ represents the system efficiency, 
and $\eta_{\text{pol}}$ is the number of polarisation states.  
A degradation factor for the noise in a continuum image is denoted as $S_{D} = 2$. 
In our work, the channel bandwidth $\Delta\nu$ is adopted as 0.3 times the frequency~\cite{Braun:2019gdo}.
On the other hand, a fixed value $\Delta\nu=300$~MHz is used in Ref.~\cite{Cembranos:2019noa,Bhattacharjee:2020phk,Dutta:2020lqc,Kar:2018rlm}.
This difference makes our $S_{\text{min}}$ larger (smaller) than others at the lower (higher) frequency regions.
The minimum flux is related to the effective collecting area $A_{e}$ 
divided by the total system noise temperature  $T_{\text{sys}}$. 
The change of the ratio $A_{e}/T_{\text{sys}}$ with frequency can be found in Fig.~7 of Ref.~\cite{Braun:2017hi}. 
Here, $\mathcal{T}$ represents the total integration time.

The first SKA phase (SKA1) can detect a minimum flux  
$3 \times 10^{-21}$~erg~cm$^{-2} s^{-1}$ for $\mathcal{T}$ equal to 100 hours at 1 GHz. 
However, for the second SKA phase (SKA2), the minimum energy flux could be lowered down to 
$\approx 3 \times 10^{-22}$~erg~cm$^{-2} s^{-1}$. 
In the following sections, we will discuss both the SKA1 and SKA2 sensitivities 
for DM annihilation cross section and decay rate.

\section{DM radio emission and energy spectrum distribution} 
\label{sec:radio}

One of most likely radio origins may be DM, 
because its annihilation or decay to the relativistic electrons and positrons produce 
the radio emission via synchrotron for $m_\chi>\mathcal{O}(100\mev)$ or 
inverse Compton (IC) for $m_\chi<\mathcal{O}(100\mev)$.  
In this section, we start with the analytical solution of the electron and positron propagation equation.
A simple assumption is that the electrons and positrons propagate with no convection, no reacceleration, 
and a very large time scale for fragmentation and radioactive decay.  
Hence, a simple propagation equation can be written as
\begin{equation}
    -\nabla\left[ D(E,\textbf{r}) \nabla \frac{\partial n_e}{\partial E}\right] 
    -\frac{\partial }{\partial E}\left[b(E,\textbf{r})\frac{\partial n_e}{\partial E}\right]
    =Q(E,\textbf{r}), 
    \label{eq:diffep}
\end{equation}
where the source term of DM annihilation/decay is 

\begin{equation}
Q_e(E,r)=
 \left\{
        \begin{array}{ll}
            \frac{1}{2}\left[\frac{\rho(r)}{m_{\chi}}\right]^2 \sv \frac{dN_e}{dE} 
            & \quad  {\rm for~DM~annihilation}, \\
            \frac{\rho(r)}{m_{\chi}}\times \frac{1}{\tau}\times \frac{dN_e}{dE} & \quad {\rm for~DM~decay.}
        \end{array}
    \right.
\label{source_function}
\end{equation}
The annihilating DM properties are described by 
the DM mass $m_\chi$, velocity averaged annihilation cross section $\sv$, normalized DM annihilation energy spectrum $\dnde$
and DM halo profile density $\rho$.  
For decaying DM, the decay time $\tau$ is inversely proportional to decay rate. 
Note that $Q_e(E,r)$ is proportional to $\left[\frac{\rho(r)}{m_{\chi}}\right]^2$ for annihilating DM 
while $\frac{\rho(r)}{m_{\chi}}$ for decaying DM.

During the propagation, the electrons and positrons interact with the interstellar medium gas, CMB, and galactic magnetic fields. 
The diffusion coefficient can be modeled as 
\begin{equation}
D(E)=D_0\times E^\gamma, 
\end{equation}
where we take $\gamma=0.3$ for the Kolmogorov description.  
According to the study of the measured B/C ratio data in the Milky Way, 
the diffusion coefficient $D_0$ ranges from 
$10^{27}$ to $10^{29}~{\rm cm}^2 s^{-1}$~\cite{Maurin:2001sj,Webber:1992dks}.

The energy loss term $b(E, \mathbf{r})$ accounts the total energy loss via 
synchrotron, IC, Coulomb, and bremsstrahlung processes. 
The explicit form of the energy loss term is 
\begin{equation}
\begin{aligned}
b(E, \mathbf{r}) &=b_{\rm IC}(E)+b_{\rm syn.}(E, \mathbf{r})+b_{\text {coul. }}(E)+b_{\text {brem. }}(E) \\
&=b_{\rm IC}^{0} E^{2}+b_{\rm syn.}^{0} B^{2}(r) E^{2} +
b_{\text {coul. }}^{0} n_{e}\left[1+\log \left(\frac{E / m_{e}}{n_{e}}\right) / 75\right]\\
& +b_{\text {brem. }}^{0} n_{e}\left[\log \left(\frac{E / m_{e}}{n_{e}}\right)+0.36\right], 
\end{aligned}
\end{equation}
where $n_e$ represents the average thermal electron density. 
In this work, we take $n_e=10^{-3} {\rm cm}^{-3}$ for A2199~\cite{Storm:2012ty} and 
$n_e=10^{-6} {\rm cm}^{-3}$ for Segue1, Draco, and DF44~\cite{Colafrancesco:2006he}. 
Following Ref.~\cite{Colafrancesco:2006he,Colafrancesco:2014coa}, 
we simply take the energy loss factors $b_{\rm IC}^{0}=0.25$, $ b_{\rm syn.}^{0}=0.0254$, 
$b_{\rm coul. }^{0}=6.13$, and $b_{\rm brem. }^{0}=1.51$ in units of $10^{-16}\gev/s$. 
Moreover, the averaged values of magnetic field of four sources are 0.0880 $\mu$G (Draco), 0.0475 $\mu $G (Segue1), 
3.57 $\mu$G (A2199), and 0.432 $\mu$G (DF44). Here, the magnetic field and plasma densities are averaged within the diffusion zone.

For the boundary condition, the positron density is generally assumed to vanish at the boundary of the diffusive zone.  
In terms of the Green's function $G\left[\textbf{r}, \lambda(E,\varepsilon)\right]$, the solution arising from  Eq. \eqref{eq:diffep} is
\begin{equation}
  \frac{\partial n_e}{\partial E}(E,\textbf{r})=\frac{1}{b(E,\textbf{r})}
  \int^{m_\chi}_{E}  d\varepsilon\,
  G\left[\textbf{r}, \lambda(E,\varepsilon,\textbf{r})\right]
  Q(\textbf{r}, \varepsilon).
  \label{eq:dndeGreen}
\end{equation}
The mean free path of $e^\pm$ is denoted as $\lambda$,  
\begin{equation}
\lambda^2(E,\varepsilon,\textbf{r})=4\int^{\varepsilon}_E \frac{D(\varepsilon')}{b(\varepsilon',\textbf{r})} d\varepsilon' 
\end{equation}
with the $e^\pm$ energy $\varepsilon'$ before the propagation.  
Note that Green's function must vanish on 
the edge of leaking box due to boundary condition. 
Therefore, we can simply apply the method of image charges to 
obtain the Green function. 
With the \textit{image charge} of each real charge, we can sum of real charges and 
all its image charges at the position $r_n=(-1)^n r+ 2 n r_h$.
Each charge can be treated as free-space Green functions,
\begin{equation}
G\left[\textbf{r}, \lambda(E,\varepsilon,\textbf{r})\right] =\sum_{n=-\infty}^\infty(-1)^n G_{\rm free}\left(\textbf{r}, \lambda\right),
\end{equation}
where a free-space Green function\footnote{Instead of using the image charge method, 
another method proposed by Ref.~\cite{Vollmann:2020gtu} shows that 
the Green function can be expanded by a Fourier series. 
We have confirmed that the maximum difference between two numerical results is around $50\%$.} is given by
\begin{equation}
G_{\rm free}\left[\textbf{r}, \lambda(E,\varepsilon,\textbf{r})\right]=\frac{1}{\sqrt{\pi \lambda^2}}
\int^{r_h}_0 dr' \frac{r'}{r_n}\left[\frac{\rho(r')}{\rho(r)}\right]^\alpha\times
\left[ e^{-\frac{(r'-r_h)^2}{\lambda^2}} 
-e^{-\frac{(r'+r_h)^2}{\lambda^2}} \right].
\label{eq:Gfree}
\end{equation}
Again, the index $\alpha$ is for either decaying DM $\alpha=1$ or annihilating DM $\alpha=2$.

\begin{figure*}[htbp]
\begin{centering}
\subfloat[]{
\includegraphics[width=0.49\textwidth]{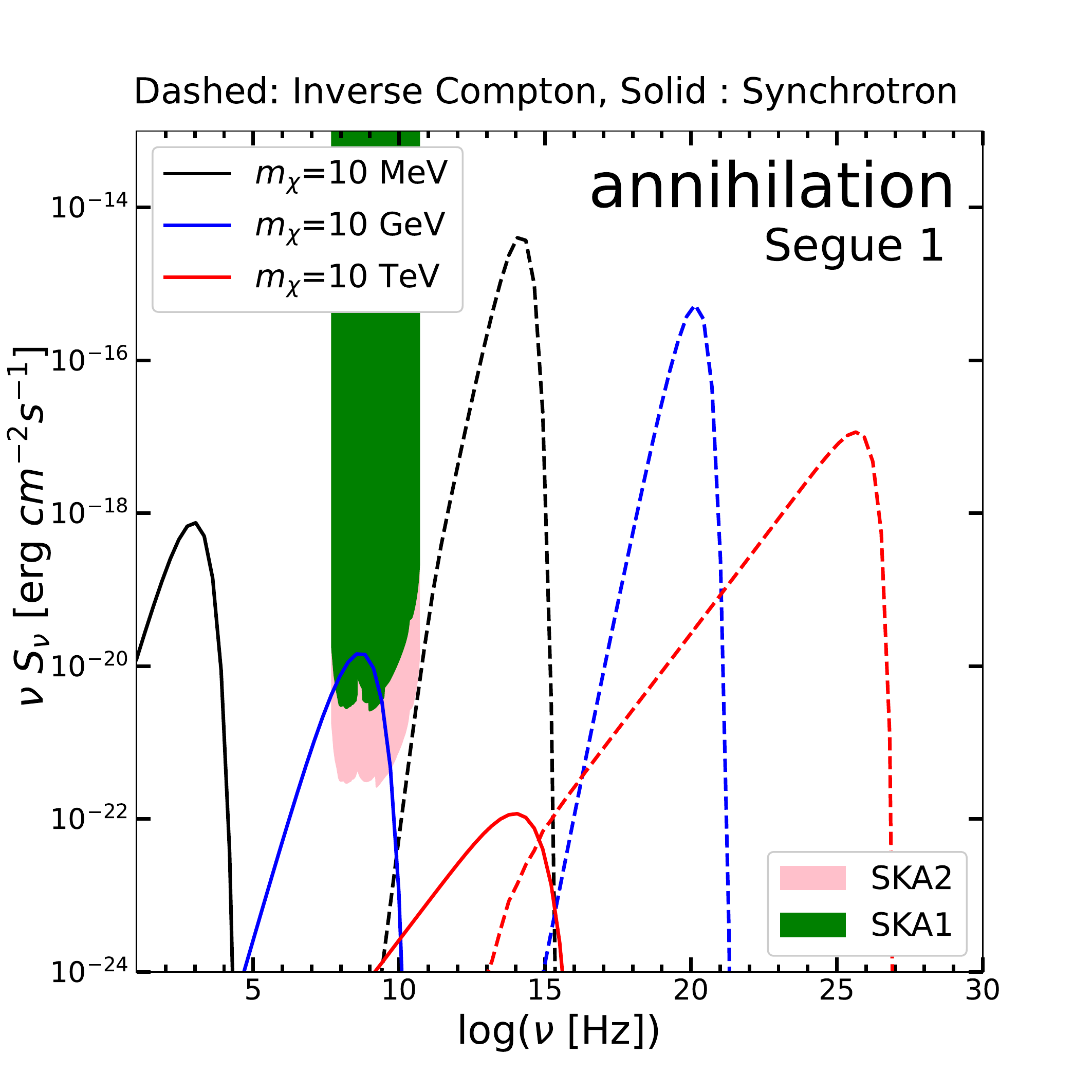}
}
\subfloat[]{
\includegraphics[width=0.49\textwidth]{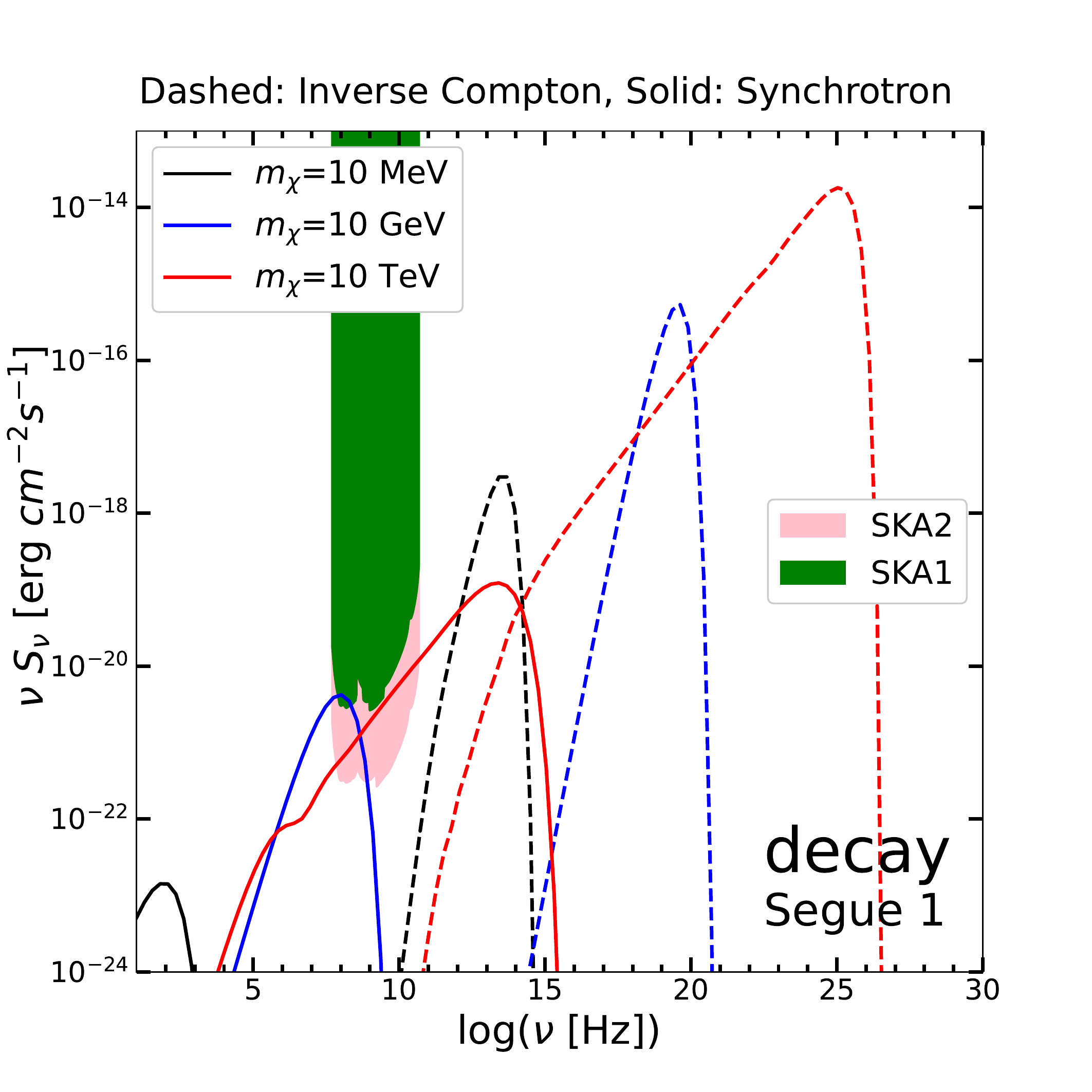}
}
\caption{
The energy spectrum distribution from Segue I.  
The $e^+e^-$ final state are considered and  
three benchmark DM masses are $10\mev$ (blue), $10\gev$ (black), and $10\tev$ (red).   
The DM annihilation cross section (left panel) 
is fixed to be $\sv=10^{-28}$ cm$^3 s^{-1}$ while the DM decay time (right panel) is $\tau=10^{28}~s$. 
For DM halo model, we hire the Einasto profile as presentation. 
}
\label{Fig:SED_mx}
\end{centering}
\end{figure*}

Plugging the green function back to Eq.~\eqref{eq:dndeGreen}, 
we can finally write the energy spectrum distribution (SED) 
for ICS and synchrotron as 
\begin{equation}
S_\nu(\nu)=\int d\Omega \int_{l.o.s.} \frac{dl}{4\pi} \int_E^{m_\chi} 2 dE\times (\mathcal{P}_{IC}+\mathcal{P}_{syn})\times \frac{\partial n_e}{\partial E}.
\label{eq:SED}
\end{equation}
The standard power of ICS $\mathcal{P}_{IC}$ and synchrotron $\mathcal{P}_{syn}$ 
can be found in Ref.~\cite{McDaniel:2017ppt}. 
We will see later that the maximum radio frequency of the synchrotron radiation is 
$\nu \approx 4.7~{\rm GHz}(E/{\rm GeV})^2(B/{\rm mG})$ while 
the IC power spectrum $\mathcal{P}_{IC}$ is insensitive by changing the magnetic field \cite{Profumo}. 
Except clusters, the magnetic field model used in this work is \cite{McDaniel:2017ppt}
\begin{eqnarray}
B(r) = B_{0} \exp(-r/r_{c}),
\label{magnetic_field}
\end{eqnarray}
where $B_0$ represents the central magnetic field strength. 
We adopt the half-light radius as the target center radius $r_c$. 
To describe cluster magnetic fields, 
we adopt the $\beta$-model 
\begin{equation}
B(r) \propto n_{\mathrm{gas}}(r)^{\eta}=B_{0}\left(1+\frac{r^{2}}{r_{\mathrm{c}}^{2}}\right)^{-(3 / 2) \beta \eta},
\label{magnetic_cluster}
\end{equation}
based on the simulation results where cluster magnetic field is a power law of  
the thermal gas density. 
The parameter $\beta$ is fitted by using X-ray data and its value is taken as $\beta = 0.655$ ~\cite{Bonafede2010,Storm:2012ty}. 
The $\beta$-model well agrees multiple Faraday Rotation Measures of clusters~\cite{Vacca2012}.  
The best fit values of $B_0$ and $\eta$ of A2199 in Faraday Rotation Measures are  
$B_0=11.7~\mu G$ and $\eta = 0.9$~\cite{Vacca2012, Storm:2012ty}. 

In Fig.~\ref{Fig:SED_mx}, we plot three different masses of the spectrum energy distribution. 
The projected sensitivity of SKA1 and SKA2 is given by green and pink contours.  
By assuming Einasto halo profile and $e^+e^-$ final state, 
the $m_\chi$ equal to $10\mev$, $10\gev$, and $10\tev$ are represented by 
black, blue, and red lines, respectively. 
Here, the DM annihilation cross section in the left panel is $\sv=10^{-28}$ cm$^3 s^{-1}$ 
while the DM decay rate in the right panel is $\tau=10^{28}~s$. 
We demonstrate the synchrotron contribution by the solid and 
the ICS one by a dashed line. 
The diffuse coefficient and magnetic strength are 
given by $D_0=3\times 10^{28}~{\rm cm}^2 s^{-1}$ and $B_0=1~\mu G $.

Briefly speaking, ICS contribution dominates at the higher frequency 
while the synchrotron one does at the lower frequency.
For annihilating DM scenario as shown in the left panel of Fig.~\ref{Fig:SED_mx}, 
one can read that only synchrotron emission with $m_\chi \gtrsim \mathcal{O}(1\gev)$ or 
ICS with $m_\chi \lesssim \mathcal{O}(100\mev)$ can be probed in the SKA energy range.  
Similarly, the decaying DM scenario in the right panel of Fig.~\ref{Fig:SED_mx} shows that 
the DM mass determines the energy distribution ranges. 
The SKA can only detect the decaying DM scenario via synchrotron (heavy $m_\chi$) or ICS (light $m_\chi$).

There is a tricky feature that appears in the SED of the decaying DM.  
Unlike the annihilating DM scenario, 
the SED of the decaying DM is not suppressed by the DM mass so that 
the heavier DM can generate a higher SED peak. 
From Eq.~\eqref{source_function}, one might find the this feature 
is counterintuitive to the number density $\rho/m_\chi$. 
On the other hand, the similar factor $(\rho/m_\chi)^2$ in the annihilating DM scenario 
determines the height of SED peak, despite a nonlinear scaling.     
Actually, the spectrum $\dnde$ in Eq.~\eqref{source_function} is also the function of DM mass 
and it may play an important role to determine the height of the SED peak. 
Considering the $e^+e^-$ final state, we find that the $10\tev$ DM can decay/annihilate 
to $e^+e^-$ via the final state radiation which 
leads to $\dnde$ for the $10\tev$ DM around $\mathcal{O}(10^{4})$ times higher 
than the one for the $10\gev$ DM. 
Such an enhanced factor is still not large enough to compensate for
the suppressed factor from the number density, $10^{-6}$  
by comparing the $10\tev$ annihilating DM to the $10\gev$ one.  
By contrast, the suppressed factor for the $10\tev$ decaying DM is only $10^{-3}$ 
which makes the SED of the $10\tev$ decaying DM larger than the one of the $10\gev$ decaying DM.

In the following subsections, we will demonstrate the impact of the astrophysical uncertainties 
on the SED of the annihilating DM.  
The impact of the magnetic strength will be discussed in Sec.~\ref{sec:B}, 
the impact of diffusion will be presented in Sec.~\ref{sec:D0}, and 
the signal strength of SED for different sources will be given in Sec.~\ref{sec:sources}.  
To demonstrate the astrophysical impacts, 
we only present our result based on the annihilating DM scenario.  
It is simply because the relative variation of signal intensity with respect to 
the magnetic field and diffusion does not much depend on the choice of DM scenario.

\subsection{Impact of the magnetic strength}
\label{sec:B}
\begin{figure*}[htbp]
\begin{centering}
\subfloat[]{
\includegraphics[width=0.49\textwidth]{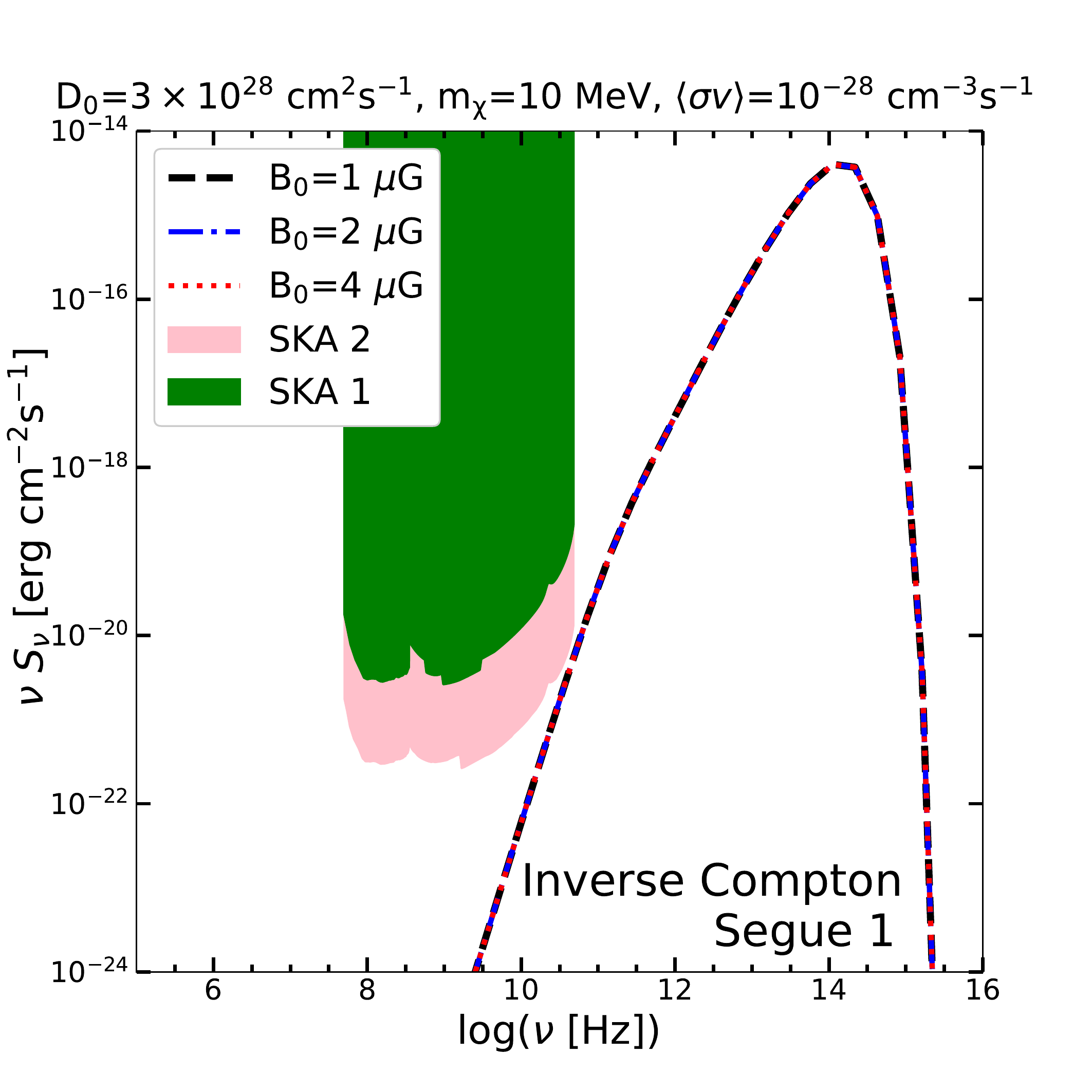}
}
\subfloat[]{
\includegraphics[width=0.49\textwidth]{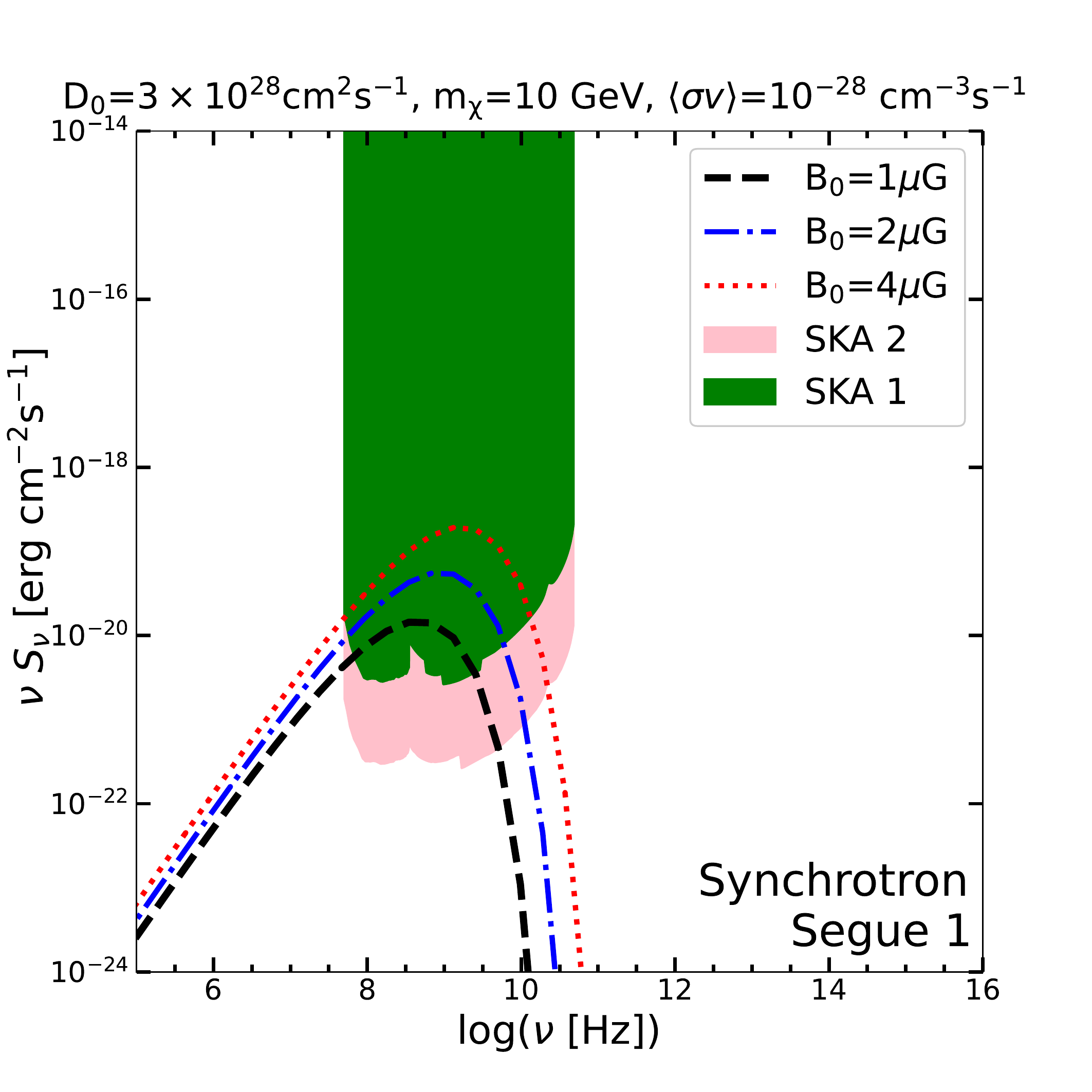}
}
\caption{  
Comparison of the SED from Segue 1 based on three different $B_0$s. 
Left panel: ICS for $m_\chi=10\mev$. 
Right panel: synchrotron for $m_\chi=10\gev$. 
Three benchmark values of $B_0$ are $1~\mu$G (black dashed line), 
$2~\mu$G (blue dashed dotted line), and $4~\mu$G (red dotted line). 
The DM is assumed to be annihilating to  $e^{+}e^{-}$. 
The annihilation cross section and the diffusion coefficient are 
taken as $\sv=10^{-28}$ cm$^3 s^{-1}$ and $D_0 = 3 \times 10^{28} \rm cm^2 s^{-1}$. 
}
\label{Fig:dif_B}
\end{centering}
\end{figure*}

The systematic uncertainties of the magnetic strength inside the galaxies remain unknown. 
We first discuss the impact of the magnetic strength on the SED and 
show the SED based on three different  $B_0$s in Fig.~\ref{Fig:dif_B} 
 for ICS (left panel) and synchrotron (right panel). 
We choose the value of the magnetic strength as $1~\mu$G, $2~\mu$G, and $4~\mu$G presented 
by black dashed line, blue dashed dotted line, and red dotted line, respectively. 
Again, we hire Einasto profile as DM halo density.  
To clearly illustrate the ICS and synchrotron, 
we show $m_\chi=10\mev$ for the ICS SED in the left panel 
but $m_\chi=10\gev$ for the synchrotron SED in the right panel.

Note that the variation of ICS with respect to the changes of $B_0$ 
is negligible but synchrotron contribution is significantly dependent on $B_0$. 
We find that three peaks of the synchrotron energy spectrum are altered non-linearly. 
This feature comes from two combined effects. 
First, the mean free path of electrons in Eq.~\eqref{eq:Gfree} 
also correlate to the value of the magnetic strength even though 
their relationship is not as simple as linear re-scaling.  
Another effect is that the observed radio peak frequency is also increasingly shifted  
if increasing $B_0$. 
Hence, changing the value of $B_0$, one can also alter the shape of the synchrotron SED.

On the other hand, the ICS SED is not sensitive to the change of the magnetic field, 
because the magnetic field strength inside dSphs is too weak to change ICS SED. 
Thus, the energy loss term ($B^2 E^2$) is not enough large to make 
a significant change of ICS SED.

In this work, we simply adopt the magnetic field with the exponential form in Eq.~\eqref{magnetic_field} and Eq.~\eqref{magnetic_cluster} 
which is also the default model in the code of \texttt{RX-DMFIT}~\cite{McDaniel:2017ppt}.
In principle, the magnetic field could be inferred from the density of the star formation rate.  
However, the extremely low density of dust and gas makes the measurement of polarization difficult 
so that there are still several choices of the magnetic field model. 
For example, the magnetic field strength inside Local Group dwarf irregulars could be 
described by a power law of the density of star formation rate, see~\cite{Chyzy:2011sw}. 
Some recent studies~\cite{Regis:2017oet,Regis:2014tga}
have reported that the center part of the magnetic field strength inside the dSphs is around $1~\mu$G.  
However, the spatial distribution of magnetic fields inside the dSphs is still 
hard to determined~\cite{McDaniel:2017ppt}.

\subsection{Impact of the diffusion coefficient}
\label{sec:D0}

\begin{figure*}[htbp]
\begin{centering}
\subfloat[]{
\includegraphics[width=0.49\textwidth]{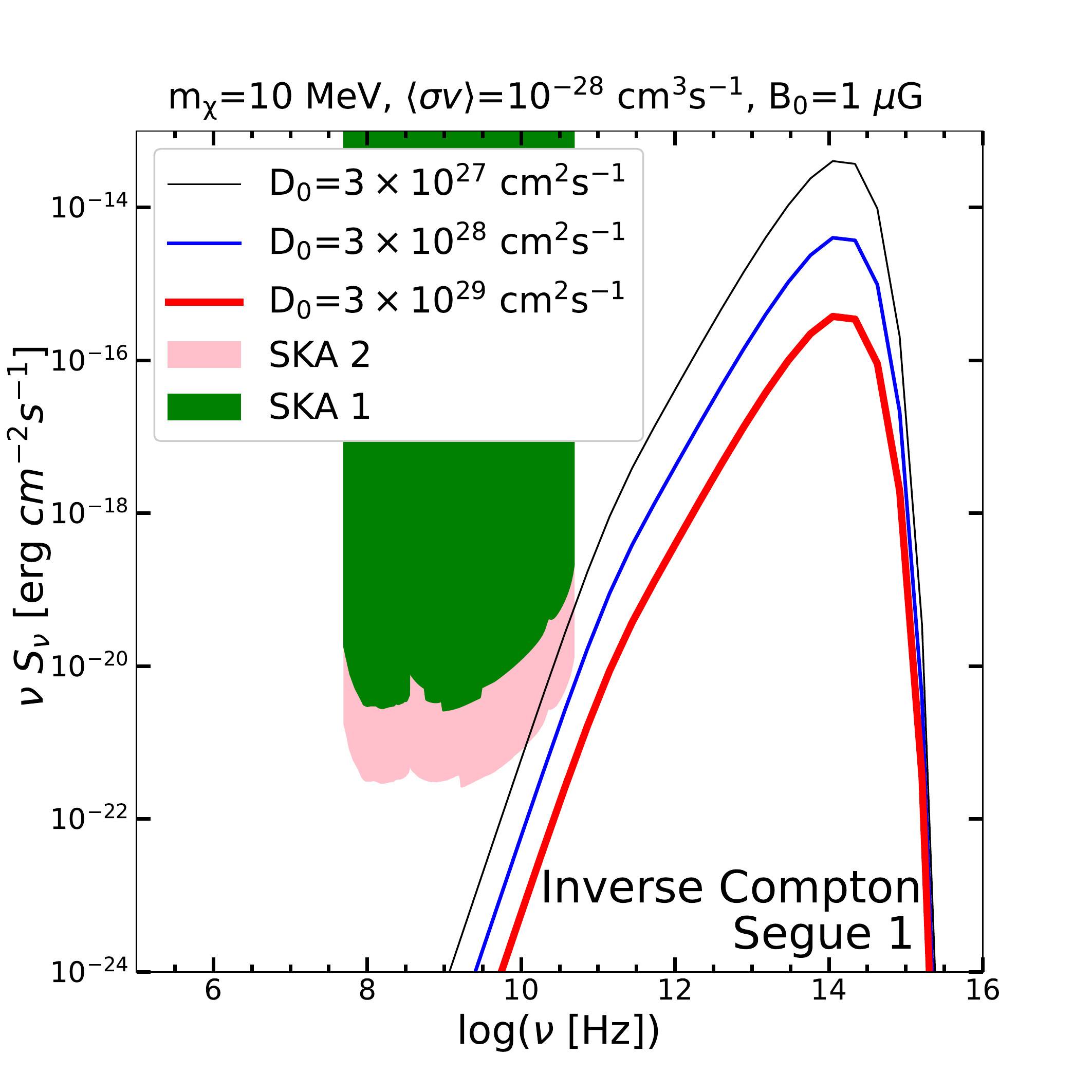}
}
\subfloat[]{
\includegraphics[width=0.49\textwidth]{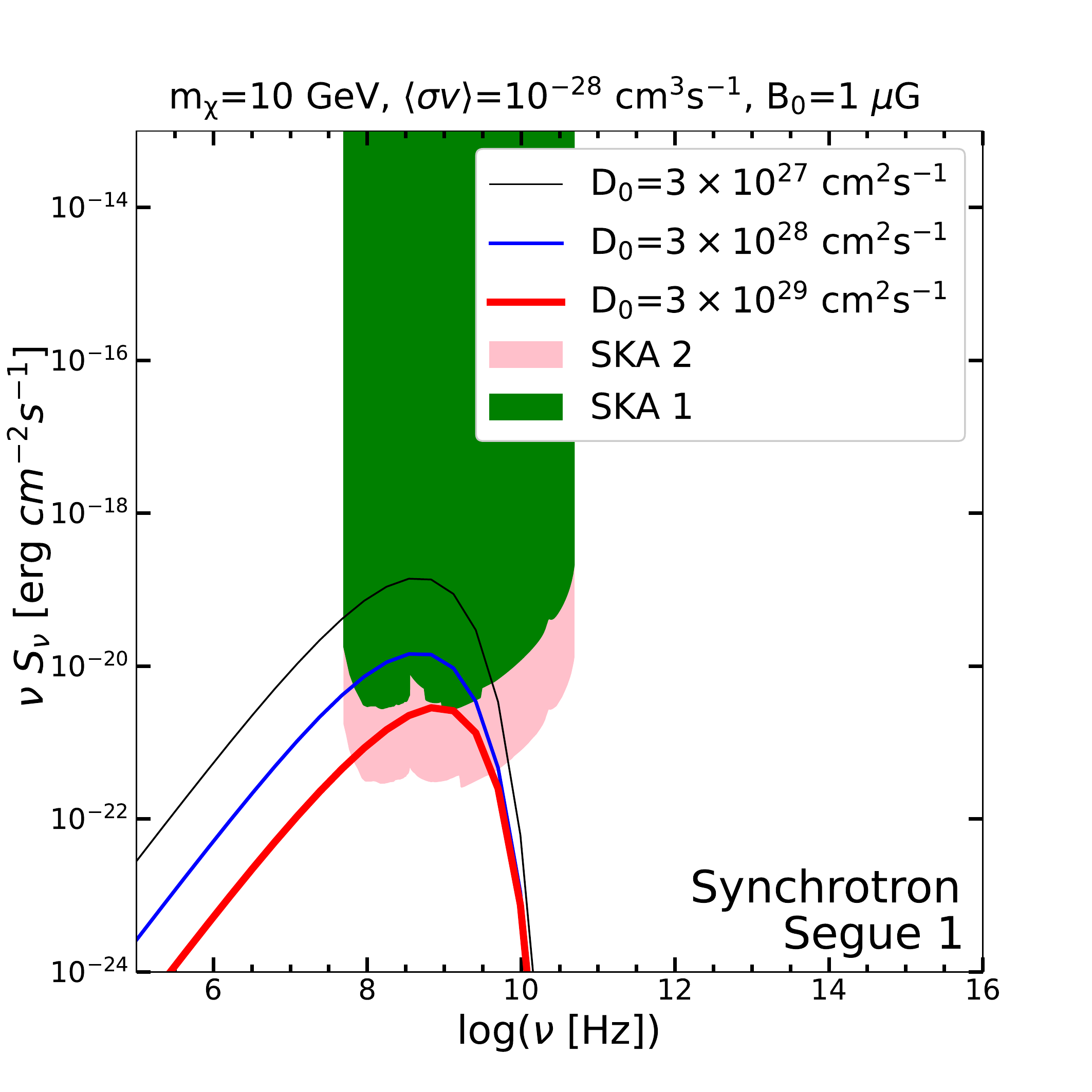}
}

\caption{
Variation of SED for Segue 1 with three different $D_0$ values. 
The annihilation cross section and magnetic field are 
fixed as $\sv=10^{-28}$ cm$^3 s^{-1}$ and $B_0=1~\mu G $, respectively. 
The left panel shows ICS for $m_\chi=10\mev$. 
The right panel shows $m_\chi=10\gev$ for the synchrotron SED. 
We adopt $e^+e^-$ as our benchmark channel. 
The red, blue and black lines denote 
$D_0 = 3 \times 10^{29} \rm cm^2 s^{-1}$, $D_0 = 3 \times 10^{28} \rm cm^2 s^{-1}$ 
and $D_0 = 3 \times 10^{27} \rm cm^2 s^{-1}$.
}
\label{Fig:dif_D}
\end{centering}
\end{figure*}

The diffusion coefficient of cosmic ray propagation is not yet precisely determined. 
In Fig.~\ref{Fig:dif_D}, we explored the SED of synchrotron and ICS with 
three possible diffusion coefficients, $D_0=3 \times 10^{27} \rm cm^2 s^{-1}$ (black), 
$D_0=3 \times 10^{28} \rm cm^2 s^{-1}$ (blue) and 
$D_0=3 \times 10^{29} \rm cm^2 s^{-1}$ (red). 
We employ the Einasto profile as the DM density distribution. 
We display ICS based on $m_\chi=10\mev$ (left panel) and synchrotron based on $m_\chi=10\gev$ (right panel). 
Here, $B_0$ is fixed to be $1~\mu$G inside Segue 1.

One can see in Fig.~\ref{Fig:dif_D} that a smaller $D_{0}$ can result in a stronger signal strength 
for both synchrotron radiation and the ICS SED.
As long as $D_{0}$ is small, the relativistic charged particles can escape from the diffusion zone 
before producing large energy loss such as synchrotron radiation via galactic magnetic fields or 
scattering with CMB photons via ICS. 
Unlike the effect that the magnetic fields determine the peak energy of the SED, 
the coefficient $D_{0}$ is only related to the height of the SED.  
As for completeness, in the result section, we will present the sensitivities of SKA1 and SKA2 by 
taking three $D_0$s together with three $B_0$s as a systematic study.

\subsection{Survey of different sources}
\label{sec:sources}

\begin{table}
 \begin{tabular}{lcccc}
  \hline    \hline
    &\qquad Draco  \cite{McDaniel:2017ppt} 
   &\qquad Segue 1  \cite{Natarajan:2015hma, Simon:2010ek}   &\qquad  A2199 \cite{Storm:2012ty,Groener:2015cxa,Babyk2014} &\qquad DF44 \cite{Wasserman:2019ttq}\\
  \hline
Distance from the Earth $l_0$   &\qquad 80~kpc    &\qquad 23~kpc &\qquad  118~Mpc &\qquad 101~Mpc\\
$r_h$ (kpc)   &\qquad 2.5    &\qquad 1.6  &\qquad 500.0  &\qquad 9.2 \\ 
$r_{\rm core}$ (kpc)  &\qquad 0.22   &\qquad 0.038  &\qquad 102  &\qquad 4.6 \\ 		
$B_0$ ($\mu$G)       & \qquad 1.0      & \qquad 1.0   & \qquad 11.7   &\qquad 1.0 \\ 			
$\rho_{s}$ ($\gev/{\rm cm}^{3}$) &\qquad 1.4   &\qquad 6.6   &\qquad 0.0854   &\qquad 0.107 \\	
$r_{s}$ (kpc)     & \qquad 1.0  &\qquad 0.15  & \qquad 340   & \qquad 9.27\\
$D_0$ (cm$^2 s^{-1}$)  &\qquad$3\times 10^{28}$  &\qquad$3\times 10^{28}$  &\qquad$3\times 10^{28}$   &\qquad$3\times 10^{28}$ \\
Angular size (deg) &\qquad 1.79  &\qquad 4.0  
&\qquad 0.24  &\qquad 0.0052  \\
Halo profile &\qquad NFW   &\qquad  Einasto  &\qquad  NFW   &\qquad NFW\\
  \hline    \hline
\end{tabular}
\caption{The preference values of model parameters for Draco, Segue 1, A2199 and Dragonfly 44 (DF44). 
 \label{table1}
}
\end{table}

Intuitively, we can choose the brightest source to set a most stringent limit 
on DM annihilation cross section $\sv$ or DM decay time $\tau$. 
However, different sources have various magnetic field strengths, 
DM halo distribution, and the distance to the Earth. 
From Eq.~\eqref{eq:SED}, we can read that those various source properties 
may constrain DM parameter space differently. 
Therefore, we compare the SED from four different sources as 
their properties summarized in table~\ref{table1}. 
We focus on two dSphs (Draco and Segue 1), one radio-poor cluster (A2199), 
and one DM rich ultra-diffuse galaxy (Dragonfly 44). 

The angular size is defined by $\alpha$ = $\arctan(r/l_0)$, 
where $l_0$ is the distance from the Earth to the target and $r$ is the target radius of region of interest.  
We evaluate the values of diffusion zone $\alpha$ for four sources are 
1.79 deg (Draco), 4.0 deg (Segue 1),  0.24 deg (A2199), and 0.0052 deg (DF44). 
The SKA has wide field-of-view (FOV) receivers and can use multiple FOVs to observe the entire sky. 
The FoV is inversely proportional to frequency-square.  
The values of FOV at different frequencies are: 
327~arcmin at 150~MHz, 120~arcmin at 300~MHz, 109~arcmin at 770~MHz, 
60~arcmin at 1.4~GHz, 12.5~arcmin at 6.7~GHz and 6.7~arcmin at 12.5~GHz ~\cite{Braun:2019gdo}. 
When driving a very conservative sensitivity, one could require $\alpha$ within a FOV. 
However, the SKA allows us to digitally patch several FOVs in order to 
cover our interesting angular size. 

Although the magnetic field strength $B_0$ in M31 can be a large value $15 \pm 3~\mu$G 
determined by the Faraday rotation measurement of polarized radio emissions, 
searching for DM induced synchrotron from M31 can be contaminated by  
the background radio of M31. Instead, we select a radio poor cluster  
A2199 as an example. 
Yet, the diffuse radio emission from A2199 is still detected as 
an upper limit at $327$~MHz by the Westerbork Northern Sky Survey~\cite{Rudnick2009}.  
Furthermore, recent observation has reported 
that Dragonfly 44 (DF44) is one of the largest ultra-diffuse galaxies (UDGs) 
in the Coma cluster and it is very dark with a high mass-to-light ratio around 
$48^{+21}_{-14}$\,M$_{\odot}$/L$_{\odot}$~\cite{vanDokkum:2016uwg}. 
Namely, around $99\%$ of total DF44 mass is DM. 
We simply fix the value of $B_0$ inside the DF44 as $1~\mu$G because 
the magnetic field in clusters can be estimated 
around $0.1-1~\mu$G based on the observation of radio halos~\cite{Carilli:2001hj}. 
Also, we can search DM radio emission from other DM rich system 
like dSphs. 
Here, we take Draco as a classic option while Segue 1 is the nearest dSph to the Earth.

For DM density distribution, we choose the NFW profile for Draco, A2199, and DF44. 
However, the NFW profile can not fit the rotation velocity data of Segue 1 well.  
Hence, we adopt Einasto profile with $\alpha=0.3$ for Segue 1~\cite{Natarajan:2015hma}. 
Regarding the diffusing coefficient, 
we simply take a typical value $D_0 = 3 \times 10^{28} \rm cm^2 s^{-1}$ for all the sources.

\begin{figure*}[htbp]
\begin{centering}
\subfloat[]{
\includegraphics[width=0.49\textwidth]{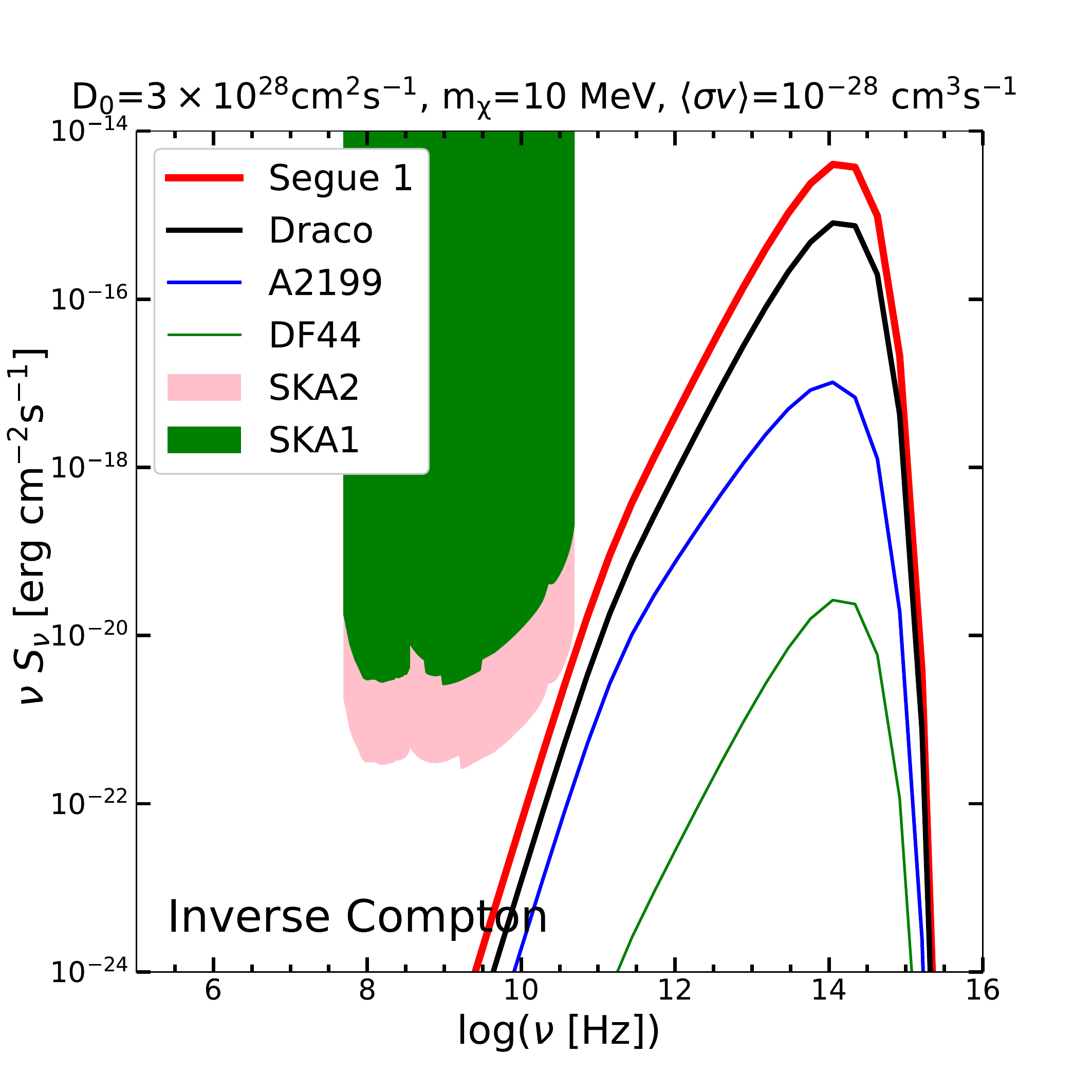}
}
\subfloat[]{
\includegraphics[width=0.49\textwidth]{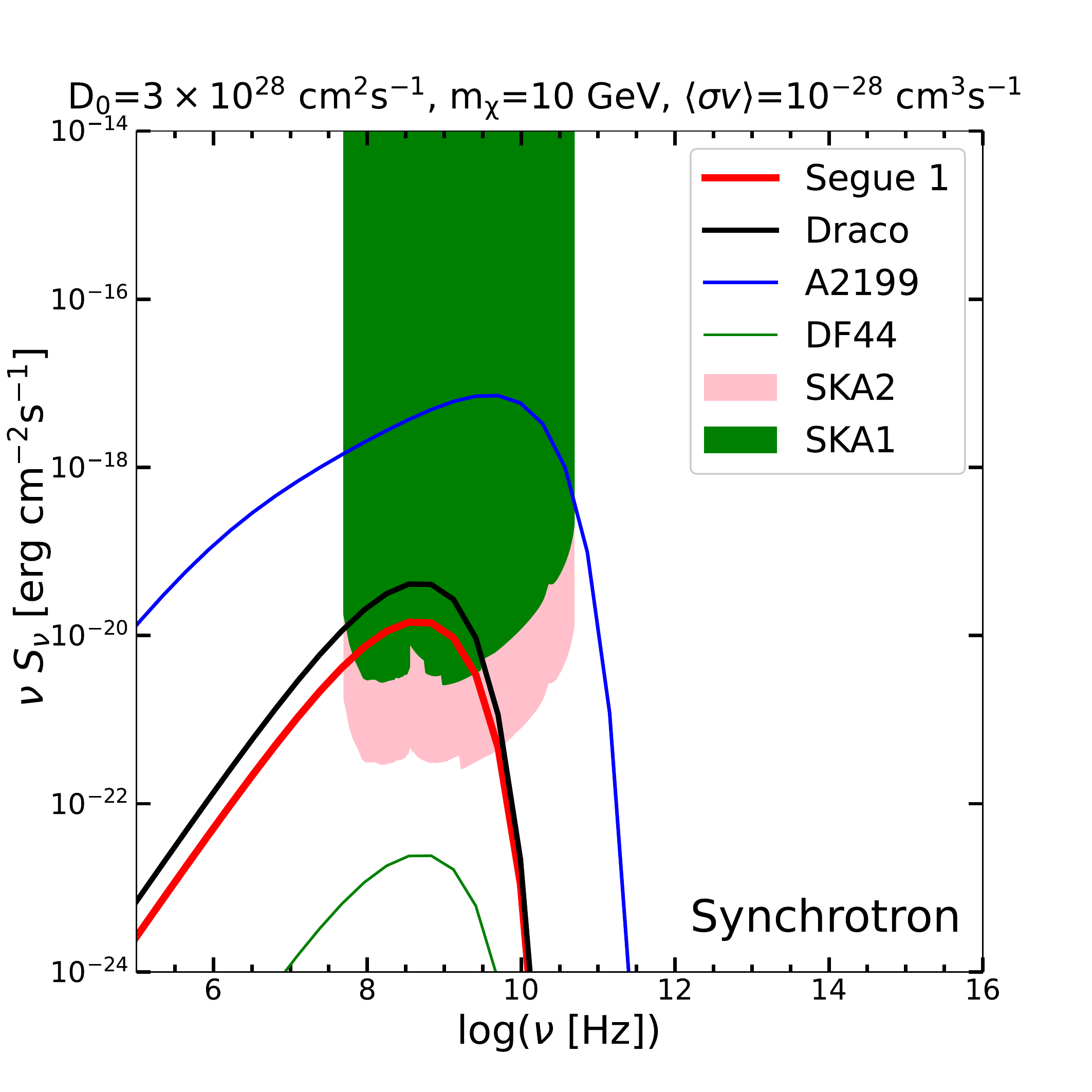}
}
\caption{ The DM energy spectra from different sources by assuming a $e^+e^-$ channel. 
The left panel is ICS SED for $m_\chi=10\mev$ while the right panel displays synchrotron 
SED for $m_\chi=10\gev$.
We compare the following four sources: Draco, Segue 1, A2199, and DF44.
}
\label{Fig:dif_source}
\end{centering}
\end{figure*}

Fig.~\ref{Fig:dif_source} demonstrates the DM SED from four different sources (Draco, Segue 1, A2199 and DF44) 
generated through ICS (left panel) and synchrotron emission (right panel). 
Again, we compute both DM induced ICS and synchrotron with 
a fixed annihilation cross section $\sv=10^{-28}$~cm$^3 s^{-1}$ 
while DM mass is $10\mev$ for ICS but $10\gev$ for synchrotron emission. 
From the right panel of Fig.~\ref{Fig:dif_source}, we found that 
DM induced ICS is basically governed by the propagation distance from the source to the Earth 
so that the predicted SED from Segue 1 is the largest 
and the one from DF44 is the smallest among these four chosen sources. 
This can be understood by the fact that ICS is not sensitive to $B_0$ as shown in Fig.~\ref{Fig:dif_B} 
but the flux is inversely proportional to the square of the distance, even if  
the size of DF44 is much larger than Segue 1. 
\textit{Hence, Segue 1 is the best source for searching DM induced ICS among these four selected sources.}

On the other hand, the DM SED generated through synchrotron emission behaves more non-trivially. 
The combined effects from the distance from the Earth, halo size, and $B_0$ make 
Segue 1 less detectable than Draco and A2199. 
The relative strength between these four sources can be partially determined 
by the distance from the Earth. 
Still, one can tell that the magnetic field $B_0$ partially determines the SED strength.  
In addition, we learn from Fig.~\ref{Fig:dif_B} that 
DM SED can widely spread within a large $B_0$ source.  
For example, the SED from A2199 is peaked at around $10^{9}$~Hz but 
the one from Segue 1 is peaked at around $3\times10^{8}$~Hz.
\textit{For detecting the DM SED generated through synchrotron, A2199 can be more promising 
than others.}

\section{Result}
\label{sec:result}

\begin{figure*}[htbp]
\begin{centering}
\subfloat[]{
\includegraphics[width=0.49\textwidth]{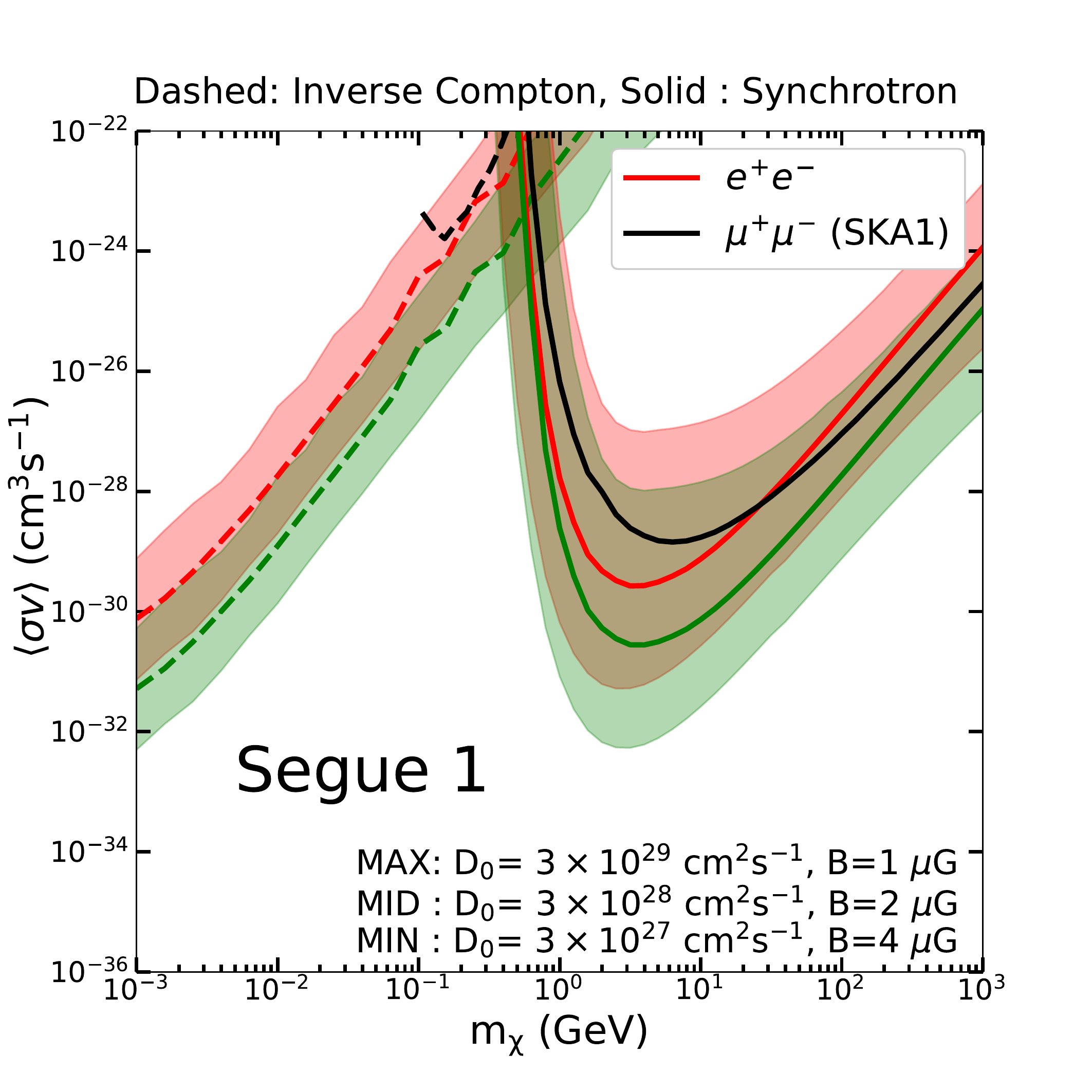}
}
\subfloat[]{
\includegraphics[width=0.49\textwidth]{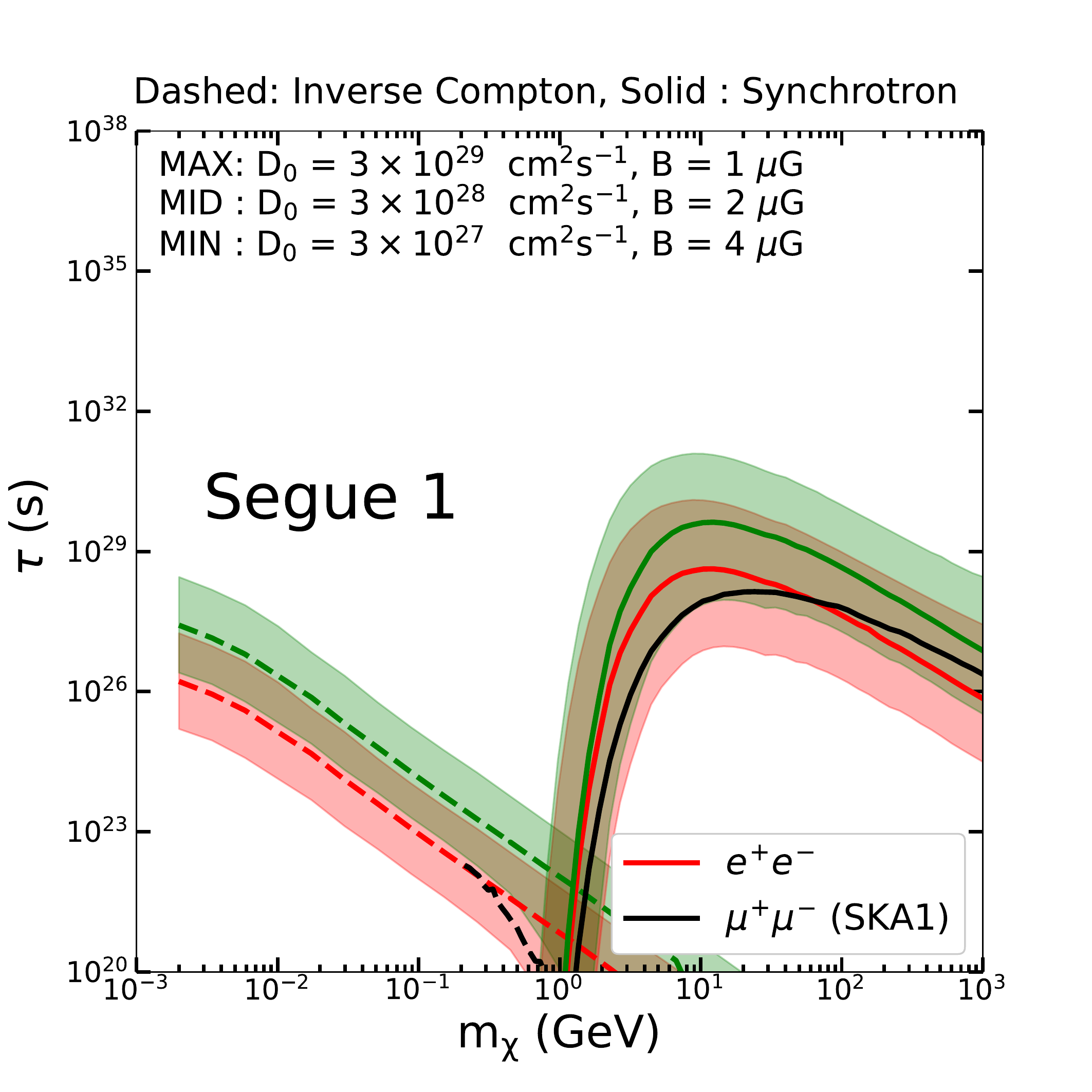}
}
\quad
\subfloat[]{
\includegraphics[width=0.49\textwidth]{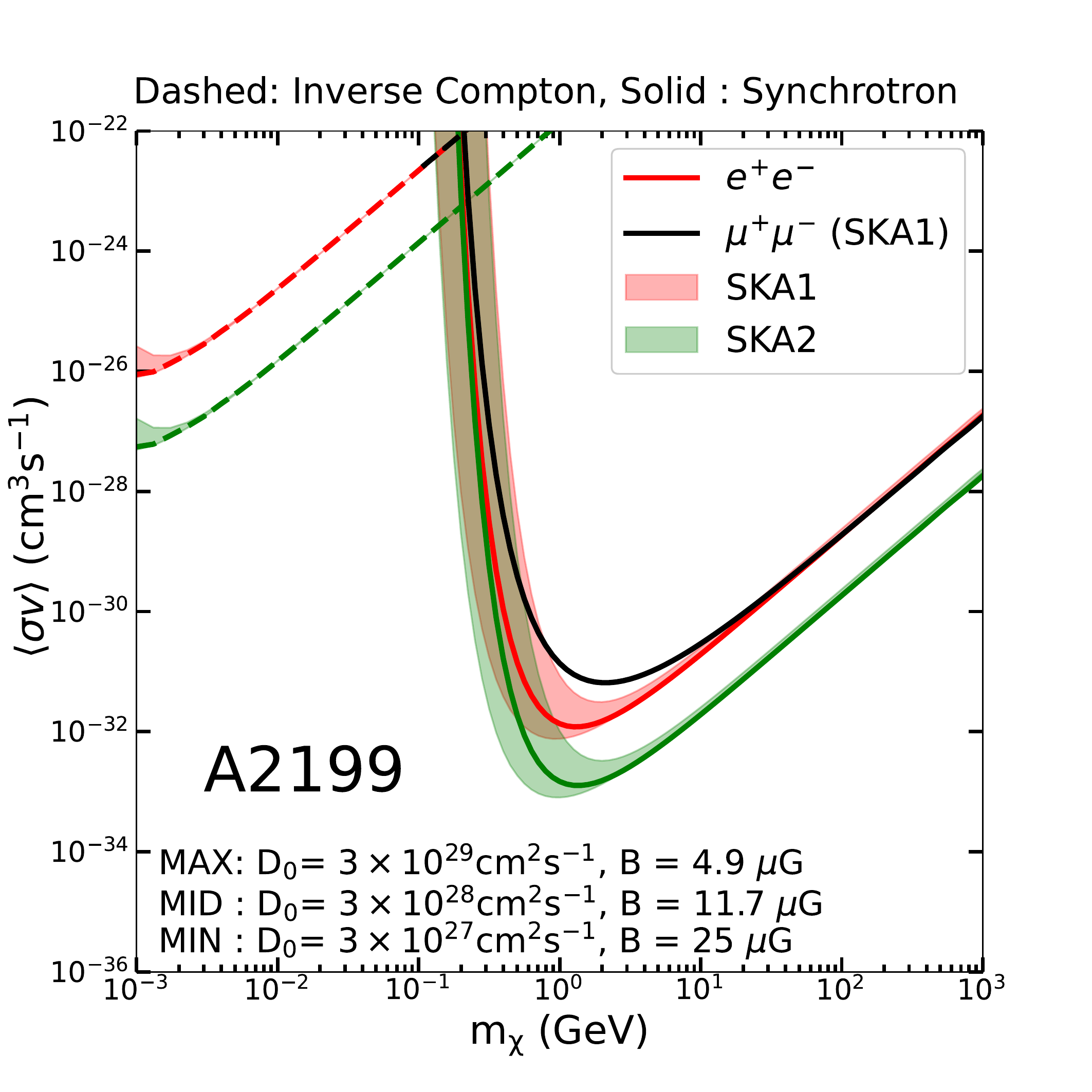}
}
\subfloat[]{
\includegraphics[width=0.49\textwidth]{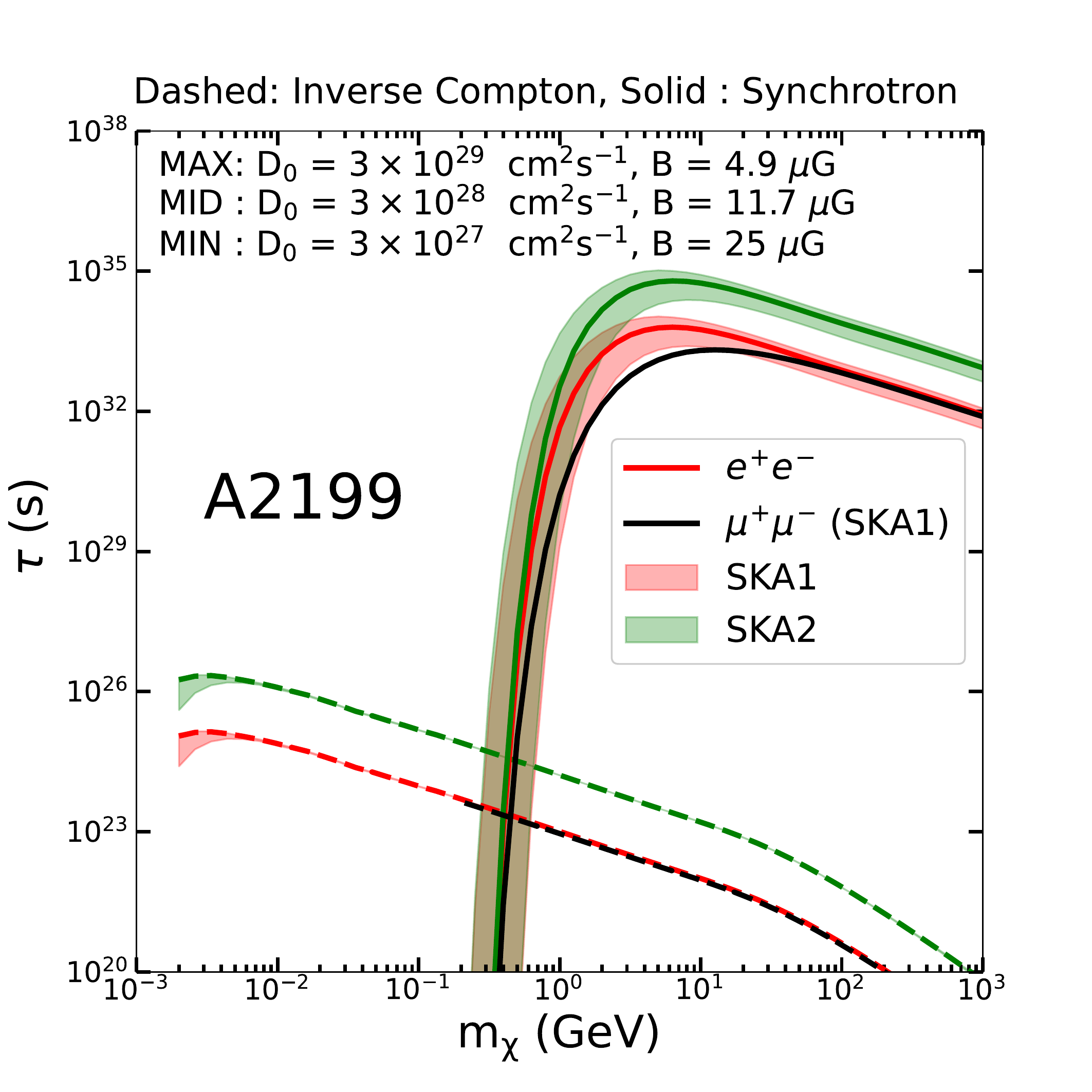}
}
\caption{
The upper limits of DM annihilation cross section $\sv$ (two left panels) and 
the lower limits DM decay time $\tau$ (two right panels) based on SKA1 (red bands) and SKA2 (green bands) 
with 100 hours of exposure.
The two upper and lower panels represent the limits from Segue 1 and A2199, respectively.  
Except for $B_0$ and $D_0$, the rest configurations of Segue and A2199 are given in table~\ref{table1}. 
The upper edge, central line, and lower edge of pink and green bands correspond to 
the configuration of \texttt{MAX}, \texttt{MID}, and \texttt{MIN}.
The dash lines show the limits from DM induced ICS and 
the solid lines denote the limits from DM induced synchrotron fluxes.
}
\label{Fig:dif_Excure}
\end{centering}
\end{figure*}

In this section, by using Eq.~\eqref{S_min}, we project the minimum SKA1 and SKA2 flux to ($m_\chi$, $\sv$) plane 
for DM annihilation scenario but to ($m_\chi$, $\tau$) for DM decay one. 
To include the systematic uncertainties, we adopt three benchmarks, 
$D_0/10^{28}/({\rm cm}^2 s^{-1})=30, 3, 0.3$. 
For magnetic field $B_0$, it depends on the selected source. 
As discussed in Sec.~\ref{sec:sources}, the most promising source for DM ICS is Segue 1 
while the one for DM synchrotron is A2199. 
Therefore, we present SKA1 and SKA2 limits based on these two sources.
Finally, we can define three benchmarks based on six combinations of $B_0$ and $D_0$,
\begin{itemize}
\item \texttt{MAX:} $D_0=3\times 10^{29}~{\rm cm}^2 s^{-1}$ and $B_0^{\rm MAX}$,
\item \texttt{MID:} $D_0=3\times 10^{28}~{\rm cm}^2 s^{-1}$ and $B_0^{\rm MID}$,
\item \texttt{MIN:} $D_0=3\times 10^{27}~{\rm cm}^2 s^{-1}$ and $B_0^{\rm MIN}$.
\end{itemize}
Here, the selected three benchmark $B_0^{\rm MAX}$, $B_0^{\rm MID}$, and $B_0^{\rm MIN}$
are $B_0/\mu{\rm G}=1.0,~2.0,~4.0$ for Segue 1 but $B_0/\mu{\rm G}=4.9,~11.7,~25$ for A2199.

In Fig.~\ref{Fig:dif_Excure}, We show our derived upper limits of $\sv$ (two panels in the left column) and 
the lower limits of $\tau$ (two panels in the right column). 
The two upper panels are based on Segue 1 but the two lower panels are based on A2199. 
We depict the sensitivities of SKA1 and SKA2 for $e^+e^-$ final state 
as a red and green band, respectively. 
The upper and lower edges correspond to \texttt{MIN} and \texttt{MAX} configuration. 
In comparison, the black solid line is the SKA1 sensitivity for $\mu^+\mu^-$ final state. 
We note that the limits based on $e^+ e^-$ and $\mu^+\mu^-$ final state only differ 
near the threshold.

For the DM annihilation scenario, the upper limits (Segue 1) are governed by 
DM induced ICS at $m_\chi\lesssim 0.4\gev$ but 
DM induced synchrotron emission at $m_\chi\gtrsim 0.4\gev$. 
However, the crossing point of two contributions is shifted to $m_\chi\approx 0.8\gev$ 
if the decaying DM is considered. 
We can see that the limits at DM mass heavier than $1\tev$ can 
be simply obtained by linear extrapolation. 
Comparing the limits from Segue 1 (two upper panels) and 
A2199 (two lower panels), the spreads of limits from these three benchmark configurations 
are roughly three orders for Segue 1 but the color bends for A2199 is much shrunk. 
This is because diffusion zone radius $r_h$ of A2199 is much larger than 
the one of Segue 1. Using the same diffusion coefficient,  
the electrons and positrons in A2199 are more likely confined within $r_h$ than in Segue 1.
Moreover, we also learn that the systematic uncertainties 
are overwhelmed by $B_0$ than the one from $D_0$. 

\begin{figure*}[htbp]
\setlength{\abovecaptionskip}{0.cm}
\begin{centering}
\subfloat[]{
\includegraphics[width=0.49\textwidth]{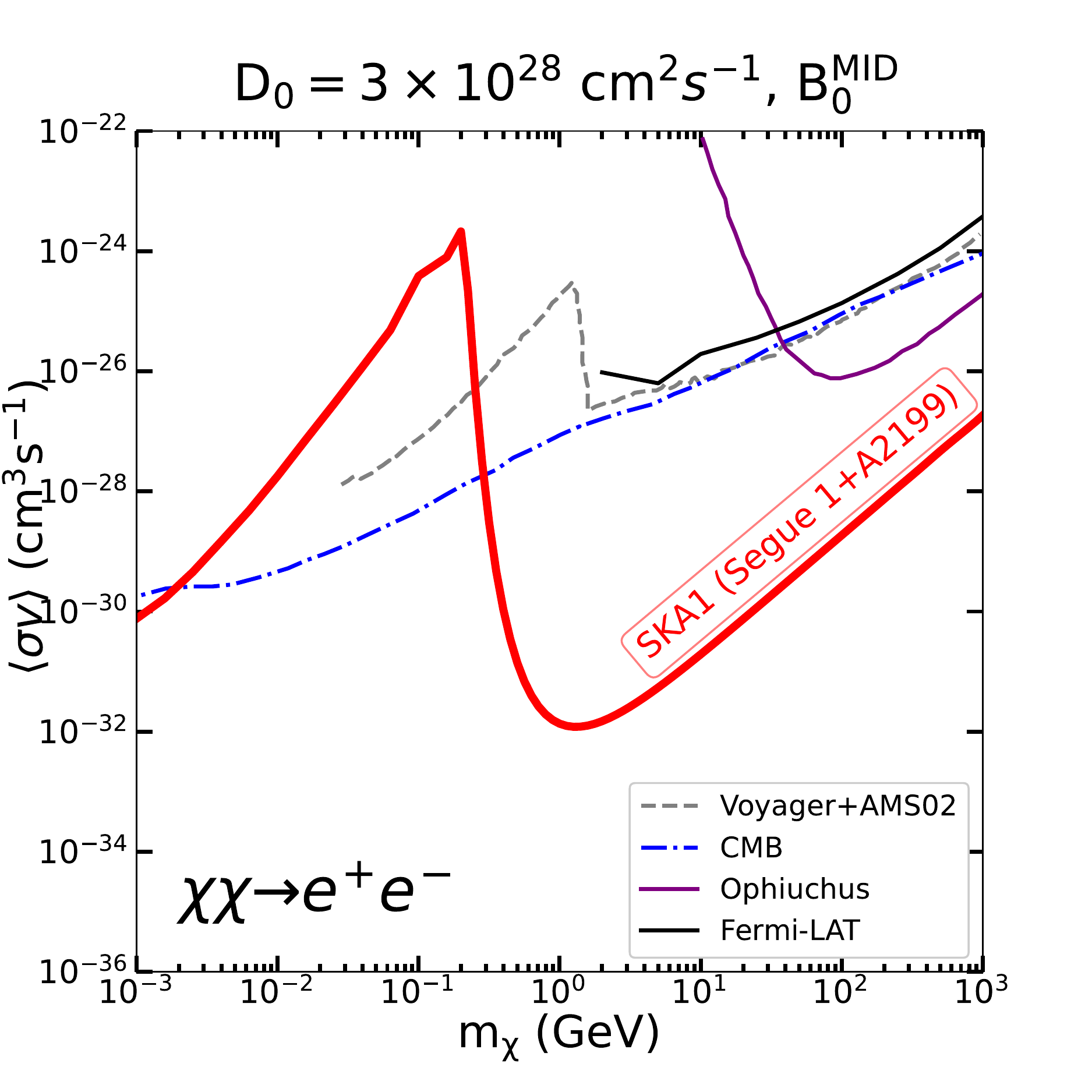}
}
\subfloat[]{
\includegraphics[width=0.49\textwidth]{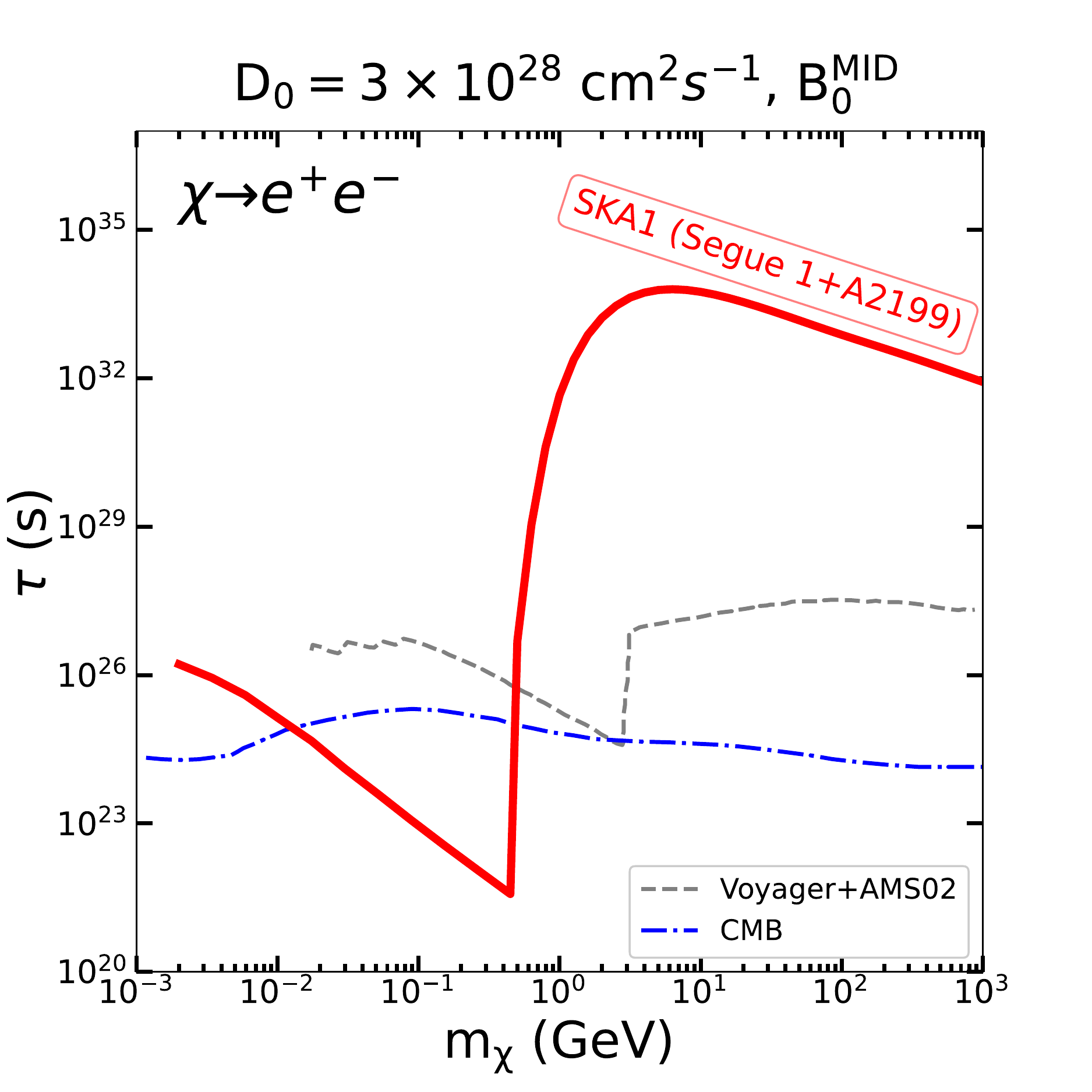}
}
\quad
\subfloat[]{
\includegraphics[width=0.49\textwidth]{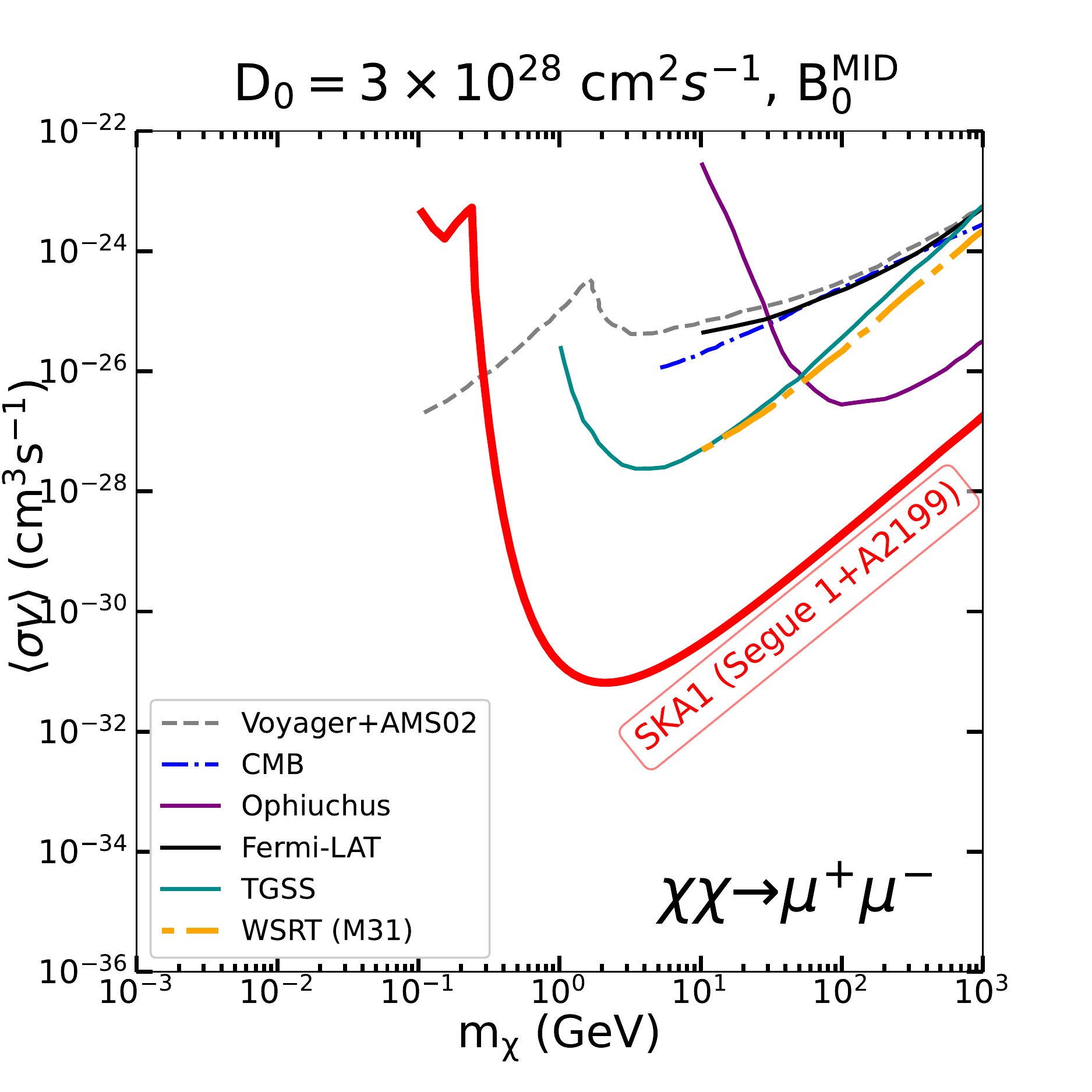}
}
\subfloat[]{
\includegraphics[width=0.49\textwidth]{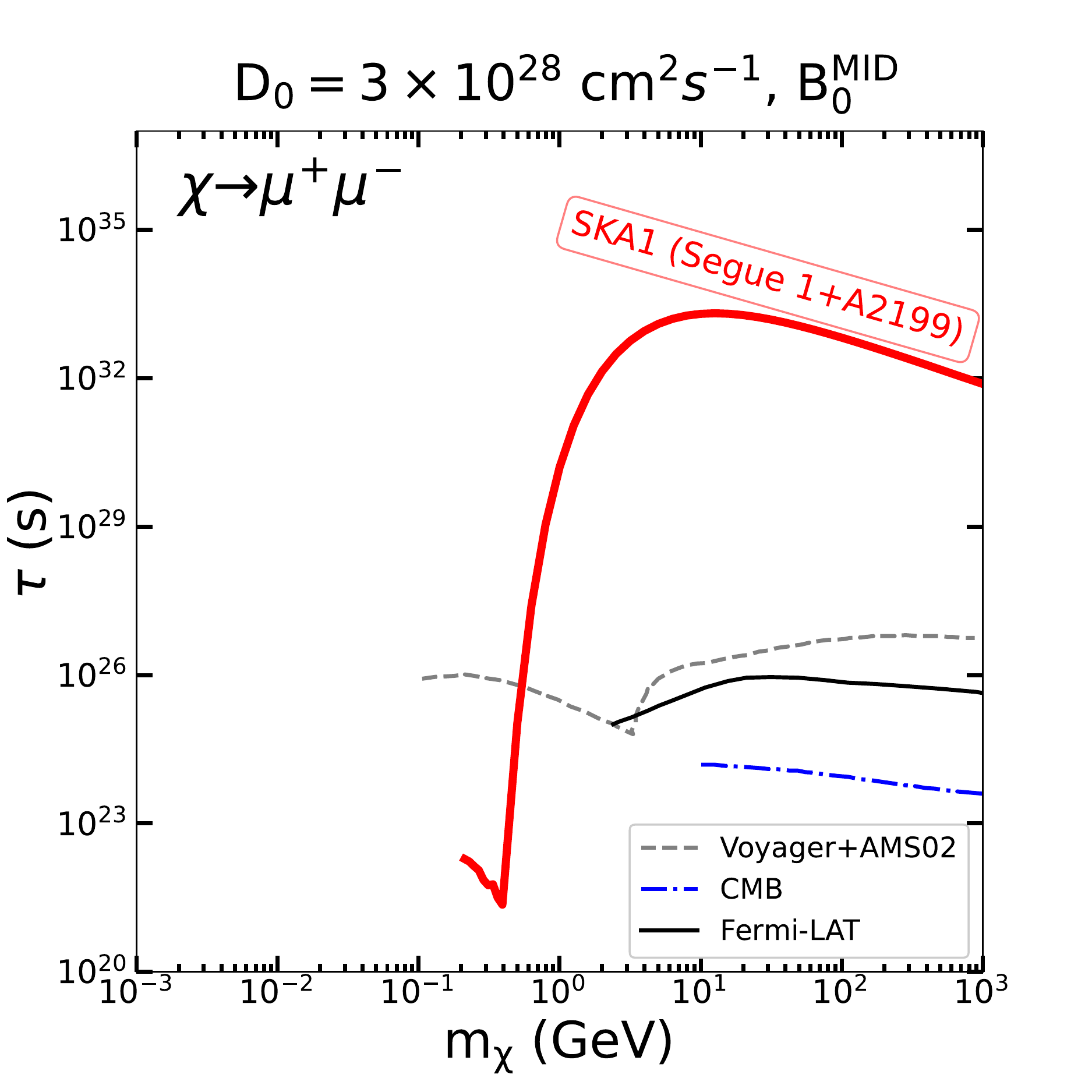}
}
\caption{Comparisons with other existing limits based on $e^+e^-$ (upper panels) and $\mu^+\mu^-$ (lower panels) final state. 
The annihilating and decaying DM scenarios are shown in the right and left panels. 
We plot the SKA1 limit with \texttt{MID} configuration as the red lines 
where we combine two limits from ICS 
at the small $m_\chi$ region of Segue 1 and 
synchrotron emission at the large $m_\chi$ region of A2199. 
}
\label{Fig:cf}
\end{centering}
\end{figure*}

By combining the result from Segue 1 and A2199, for $e^+e^-$ final state 
we except the future SKA1 can probe the region $\sv\gtrsim 10^{-32}$~cm$^3 s^{-1}$ 
for annihilating DM and $\tau \gtrsim 10^{34}~s$ for decaying DM at $m_\chi\simeq 1\gev$ . 
For $\mu^+ \mu^-$ final state, the limits are weaker about one order magnitude than the former.
The limits will be further improved by SKA2 by around one order magnitudes. 
Strikingly, this is a piece of good news for the annihilating DM scenario
because the SKA may offer a chance to detect a very low DM annihilation 
cross section in the present universe, 
for example DM annihilation only via $p$-wave or off-resonance at the present.

Finally, we compare our results with other previous existing limits in Fig.~\ref{Fig:cf}
by choosing the $e^+e^-$ (upper panels) and  $\mu^+\mu^-$ (lower panels) final state.  
The limits of Ophiuchus from VLA are shown by purple lines~\cite{Murgia2010,Beck2020}.
The stacking analysis of 23 dSph galaxies~\cite{Basu:2021zfg} is indicated by darkcyan solid lines (TGSS). 
The orange long dashed line is taken from~\cite{Chan:2016wpu} 
which uses the M31 result from the Westerbork Synthesis Radio Telescope (WSRT)~\cite{Giessubel:2013via}. The limits obtained by CMB data and Voyager and AMS02 data are shown by blue dash-dotted lines 
and gray dash lines ~\cite{Cang:2020exa,Slatyer:2016qyl,Boudaud:2016mos}.
The limit of Fermi LAT~\cite{Baring:2015sza,Ackermann:2015zua} is also given by thin black lines.  
Here, we show SKA1 limits by combining the result from DM induced ICS and synchrotron emission. 
The limit for the large $m_\chi$ region is obtained by using DM induced synchrotron emission  
with the source configuration Segue 1 \texttt{MID}.  
On the other hand, the limit for the small $m_\chi$ region is based on 
the ICS one with the source configuration A2199 \texttt{MID}. 
The crossing point between the large and small $m_\chi$ region 
is at $m_\chi\approx 0.2\gev$ for annihilating DM but $m_\chi\approx 0.4\gev$ for decaying DM. 
We can see that the current limits from dSphs are approximately four order magnitude (annihilating DM) 
and six order magnitude (decaying DM) weaker than   
the future SKA1 sensitivity at the heavy DM mass region.  
Our limits are also two orders of magnitude stronger than VLA Ophiuchus limits. 
However, if only considering Segue 1 as a comparison, SKA1 limits could be stronger than current dSphs limits 
by one order magnitude for both DM scenarios.
For the region near the detector threshold $m_\chi<10\gev$, 
the SKA shows its capability in detecting the low mass region.

\section{Summary}
\label{sec:conclusion}
In this work, we systematically study the future SKA detectability of MeV to TeV DM annihilation 
or decay signal. 
The detected frequency of radio in the SKA is between 50~MHz to 50~GHz.   
Within the required range, $e^+e^-$ can emit radio depending on their energies.   
First, the relativistic $e^+e^-$ whose kinetic energies are larger than $\mathcal{O}(100\mev)$ 
can emit radio via synchrotron with the galactic magnetic field. 
On the other hand, a soft $e^+e^-$ spectrum with kinetic energies less than $\mathcal{O}(100\mev)$ 
can only produce radio via ICS with CMB photons. 
Thus, the radio emission falling to the SKA detected frequency range generated via DM induced ICS or synchrotron 
requires DM mass lighter or heavier than $\mathcal{O}(100-200)\mev$, respectively. 
To set an upper limit on DM annihilation cross section or a lower limit on decay time, 
we consider both SED of DM induced ICS and synchrotron 
by comparing with the minimum flux of two SKA phases (SKA1 and SKA2).      
 
We compute the both SED of DM induced ICS and synchrotron based on several selected values of 
magnetic field strength $B_0$ and diffusion coefficient $D_0$. 
By using $m_\chi=10\mev$ for ICS and $m_\chi=10\gev$ for synchrotron, 
we again confirm that the impact of different $B_0$ on SED is more important for synchrotron than ICS. 
When changing to a larger value of $B_0$, not only the SED of DM induced synchrotron can be enhanced 
but also its peak is shifted to a larger frequency.  
However, changing to a larger value of $B_0$ only alters the SED of DM induced ICS insignificantly. 
Moreover, if varying the value of $D_0$, we note that both ICS and synchrotron SED are also modified accordingly.

We then attempt to figure out which source could be the best target for DM search based on 
the SKA sensitivity and frequency range. 
To perform a comparison, we select four different sources: 
Draco, Segue 1, A2199, and DF44.  
Here, A2199 is a radio-poor cluster, DF44 is a DM-rich ultra-diffuse galaxy, 
but Draco and Segue 1 are dSphs. 
Since DM induced ICS does not sensitive to $B_0$, the closer source is better for the SKA observation. 
Therefore, we found Segue 1 is the best source of searching for MeV scale DM among these four sources. 
However, DM induced synchrotron is more complicated because of the combined effects from different $B_0$ and 
the distance from the Earth. 
In the end, we conclude that the A2199 is a more promising source than others 
for $m_\chi\gtrsim\mathcal{O}(\gev)$ because of its largest $B_0$ even if it is further than others.

After presenting the SED changes based on different values of $B_0$ and $D_0$ as well as 
using different sources, we finally select three different benchmark configurations of 
$B_0$ and $D_0$. Based on these three representative configurations (\texttt{MAX}, \texttt{MID}, and \texttt{MIN}), 
we project the sensitivity of SKA1 and SKA2 to ($m_\chi$, $\sv$) for annihilating DM scenario and 
($m_\chi$, $\tau$) for decaying DM scenario by choosing $e^+e^-$ and $\mu^+\mu^-$ final states 
as demonstration. 
From MeV to TeV DM mass, the most stringent limit appears at around GeV where DM induced synchrotron 
still dominates at the SKA detected frequency range.

Comparing with other studies of DM radio probes with the future SKA sensitivity, 
our work includes three new differences.     
First, we compute the minimum fluxes by using a frequency-dependent bandwidth. 
The resulted limits on DM interaction can be slightly stronger than the ones used 
a frequency-independent bandwidth.  
Second, we study DM induced radio signals including ICS and synchrotron emission. 
By utilizing both DM contributions, we calculate the SKA sensitivity at the DM mass gap between $100\mev$ to $5\gev$. 
Third, it is the first time to study DM rich ultra-diffuse galaxy DF44 with SKA sensitivity. 
Our result shows that the DM SED from DF44 is the lowest among the four selected sources.

We also compared the projected SKA1 limits with 
other current limits of annihilation cross section and decay time in Fig.~\ref{Fig:cf}.   
We found that the SKA projected limit is much stronger than 
other previous exist limits from various DM indirect detectors, even only considering the SKA phase 1 sensitivity. 
At the heavy DM mass region where synchrotron contribution is dominant, 
the current most stringent limits such as TGSS at $m_\chi\sim 1\gev$ ($\mu^+\mu^-$) and  
VLA Ophiuchus limit at $m_\chi\sim 1\tev$ ($e^+e^-$ and $\mu^+\mu^-$) are both weaker than the SKA1 sensitivity. 
Taking the $e^+e^-$ final state as example, the SKA1 limit shows that  
$\sv$ can be as lower as $10^{-32}$~cm$^3 s^{-1}$ 
and $\tau$ can be $10^{34}~s$ at $m_\chi=1\gev$ but 
they become $\sv\lesssim 10^{-27}$~cm$^3 s^{-1}$ and $\tau\gtrsim 10^{32}~s$ at $m_\chi=1\tev$, 
based on the A2199 \texttt{MID} configuration. 
For $m_\chi= 5\mev$, the SKA1 limits derived from ICS inside Segue 1 with \texttt{MID} configuration 
are $\sv\lesssim 10^{-29}$~cm$^3 s^{-1}$ and $\tau\gtrsim 10^{26}~s$.
If the final state is $e^+e^-$, the present CMB can set a very stringent limit in the low DM mass region 
but the SKA1 projected limit from DM induced ICS can be still stronger at the region $m_\chi\lesssim 10\mev$.

In summary, we compare the SKA1 sensitivity with current existing limits as shown 
in Fig.~\ref{Fig:cf}. 
The limits of $\sv$ and $\tau$ can be significantly improved by 
the future SKA1 and this will help us to probe a large of the DM parameter space. 
For DM with velocity dependent annihilation cross section, 
e.g. annihilation only with $p-$wave suppressed cross section or 
annihilation via resonance in the early universe, 
the SKA would play a leading role to explore such DM models.


\section*{Acknowledgments}
We thank an anonymous referee for useful comments on SKA sensitivity.
Zhanfang Chen would like to thank Xu Pan for his valuable comments and discussions. 
Q. Yuan is supported by the National Natural Science Foundation of China under Grant No. U1738205, Chinese Academy of Sciences, and the Program for Innovative Talents and Entrepreneur in Jiangsu.

\end{document}